\documentclass[aps,graphicx,twocolumn]{revtex4}
\usepackage{amssymb}
\usepackage{amsmath}
\usepackage{dcolumn}
\usepackage{graphicx}
\usepackage{mathrsfs}
\begin{document}


\title{Faithful entanglement purification for high-capacity quantum communication
with two-photon four-qubit systems\footnote{Published in Physical Review Applied \textbf{10}, 054058 (2018)} }

\author{Guan-Yu Wang,$^{1,+}$
Tao Li,$^{2,3,}$\footnote{The first two authors contributed
equally to this work.} Qing Ai,$^{1,3}$ Ahmed Alsaedi,$^{4}$
Tasawar Hayat,$^{4,5}$  and Fu-Guo Deng$^{1,}$\footnote{Corresponding author: fgdeng@bnu.edu.cn}
}

\address{$^{1}$ Department of Physics, Applied Optics Beijing Area Major Laboratory,
Beijing Normal University, Beijing 100875, China\\
$^{2}$ School of Science, Nanjing University of Science and Technology, Nanjing 210094, China\\
$^{3}$ Theoretical Quantum Physics Laboratory, RIKEN Cluster for Pioneering Research, Wako-shi, Saitama 351-0198, Japan\\
$^{4}$ NAAM-Research Group, Department of Mathematics, Faculty of
Science, King Abdulaziz University,  Jeddah 21589, Saudi Arabia\\
$^{5}$ Department of Mathematics, Quaid-I-Azam University, Islamabad
44000, Pakistan}

\date{\today }

\begin{abstract}
As the hyperentanglement of photon systems presents lots of unique opportunities in  high-capacity quantum networking, the hyperentanglement purification protocol (hyper-EPP) becomes a vital project work and the quality of its accomplishment attracts much attention recently. Here we present the first theoretical scheme of faithful hyper-EPP for nonlocal two-photon systems in two degrees of freedom (DOFs) by constructing several fidelity-robust quantum circuits for hyper-encoded photons. With this faithful hyper-EPP, the bit-flip errors in both the polarization and spatial-mode DOFs can be efficiently corrected and the maximal
hyperentanglement in two DOFs could be in principle achieved by performing the hyper-EPP  multiple rounds. Moreover, the fidelity-robust quantum circuits, parity-check quantum nondemolition detectors, and SWAP gates make this hyper-EPP works faithfully as the errors coming from practical scattering, in these quantum circuits, are converted into a detectable failure rather than infidelity. Furthermore, this hyper-EPP can be directly extended to purify photon systems entangled in single polarization or spatial-mode DOF and that hyperentangled in polarization and multiple-spatial-mode DOFs.
\end{abstract}
\maketitle

\section{Introduction} \label{sec1}

Quantum entanglement is a unique phenomenon in quantum physics and constitutes a core
resource in quantum communication networks, such as quantum teleportation
\cite{tele1}, quantum dense coding \cite{dense1,super2}, quantum key
distribution (QKD) \cite{BBM92,Ekert}, quantum secret sharing
\cite{QSS1}, quantum secure direct communication (QSDC)
\cite{QSDC1,twostep,twostepexp}, and so on. The high-quality entanglement between distant quantum nodes constitutes an efficient basis for quantum communication.
For example, two parties in a quantum-communication network can
generate a private key through the maximally entangled channel with QKD
\cite{BBM92,Ekert}, and they can also directly exchange their private message securely
using  QSDC protocols \cite{QSDC1,twostep,twostepexp}.
Hyperentanglement is an interesting phenomenon in which a quantum
system is simultaneously entangled in multiple degrees of
freedoms (DOFs) \cite{hyper0}, and it nowadays  opens up the possibility for new kinds of quantum-information processing. For example, it can be used to largely improve the channel capacity and the security of quantum communication in linear photonic superdense coding \cite{super1},
complete the deterministic Bell-state analysis \cite{BSA1,BSA2,BSA3,BSA4}, and assist deterministic entanglement purification \cite{EPP6,EPP7,EPP8,EPP9,EPP10}. In addition, the hyperentanglement can be used to implement basic quantum-computation algorithms in the one-way model \cite{cluster,oneway1,oneway2,oneway3} and construct hyperparallel quantum gates with full potential of parallel computing capability \cite{hyperQC1,hyperQC2,hyperQC3,hyperQC4,hyperQC5,hyperQC6,hyperQC7}.
So far, hyperentangled states have been demonstrated recently using different photonic DOFs \cite{hyper1,hyper2,hyper3,hyper4,hyper5,hyper6,hyper7,hyper8}.

In practice, quantum entanglement is created locally and the distribution of
quantum entanglement is a prerequisite for a quantum-communication network. However, the interaction between the photons and their environment will inevitably decrease the photonic entanglement
or even turn the photon systems into mixed states. This will significantly decrease the security and the efficiency of quantum communication. In order to establish a large-scale quantum network, the quantum repeater is proposed to suppress the decoherence caused by environmental noise \cite{repeater1,repeater2,repeater3}.
Entanglement purification is one of the essential building blocks for quantum repeaters, and it can distil a subset of high-fidelity entangled quantum systems from a set of less-entangled ones~\cite{EPP2017}.
Since the first entanglement purification protocol (EPP) was proposed with quantum controlled NOT gates and  bilateral rotations \cite{EPP1}, some practical and interesting EPPs have been presented with different methods, including conventional EPPs
\cite{EPP2,EPP3,EPP4,EPP5} and the deterministic EPPs
\cite{EPP6,EPP7,EPP8,EPP9,EPP10}. These protocols are efficient for distilling entanglement in a single DOF, such as the polarization DOF. When it comes to  high-capacity quantum communications
based on hyperentanglement, the parties in quantum communication have to distil the high-fidelity hyperentanglement from the less-hyperentangled ones, polluted by the channel noise during the entanglement distribution process. In 2013, Ren and Deng \cite{hyEPP1} proposed a preliminary hyperentanglement
purification protocol (hyper-EPP) with parity-check quantum nondemolition nondemolitions (QNDs) for two-photon four-qubit systems. Subsequently, Ren \emph{et al.} \cite{hyEPP2} largely improved the efficiency of the hyper-EPP using the quantum-state-joining method. Recently, a general hyper-EPP a for two-photon six-qubit system was proposed with parity-check QNDs and SWAP gates  \cite{hyEPP3}, and it demonstrates a universal method for distilling the hyperentanglement in the polarization and
multiple-longitudinal-momentum DOFs.

In EPPs and hyper-EPPs, the parity-check QND completes the first step by picking out the components contributing to the final entanglement, and then the SWAP gate will be used to increase the efficiency of hyper-EPPs by transferring the entanglement in a controllable way. In general, both the parity-check QND and SWAP gate for hyper-encoded photons can be constructed with the interaction between a single photon and an artificial atom embedded in a cavity. Although the fidelities and efficiencies of these two processes can, in principle, approach unity, the corresponding performance with the practical scattering degrades drastically, due to the deviation from the ideal scattering condition, such as  the nonzero side leakage of the cavity and the finite coupling strength between the cavity and the atom \cite{hyEPP1,hyEPP2,hyEPP3}. This will inevitably degrade the performance of the hyper-EPP for photon systems. In detail, the parity-check QND with unity fidelity will project the target photons into a subspace with a deterministic parity without any distortion on the photon systems. However, a nonunity-fidelity parity-check QND will introduce some error components with the orthogonal parity, which will inevitably cause errors in hyper-EPPs. Similarly, the unity-fidelity SWAP gate swaps quantum states between two photons in a perfect way, while the imperfect SWAP gate will complete the swapping process imperfectly and distort the state of the photons. These undesirable side effects can be referred to as the secondary pollution to the original photon systems. Generally speaking, all these nonideal processes will degrade the quality of the practical quantum-information-processing tasks, such as the hyper-EPP \cite{hyEPP1,hyEPP2,hyEPP3} and the hyperparallel optical computing \cite{hyperQC1,hyperQC2,hyperQC3,hyperQC4,hyperQC5}. For the hyper-EPPs, it will consume more quantum resources and also limit the reliability of the hyper-EPPs, when only nonunity-fidelity quantum processes are available. Therefore, it is of great importance to construct the faithful hyper-EPP with fidelity-robust parity-check QNDs and SWAP gates when realistic scattering conditions are involved.

In this paper, we propose a faithful hyper-EPP for two-photon systems hyperentangled in both the polarization and spatial-mode DOFs. This hyper-EPP consists of two purification steps. The first one resorts to fidelity-robust parity-check QNDs, which will pick out the effective photon pairs with higher fidelities in either the polarization or the spatial-mode DOFs. This hyper-EPP is completed directly if the purified photon pairs are of higher fidelities in both DOFs. However, the parties need to perform the second purification step to pump the higher-fidelity entanglement of different photon pairs into one photon pair by using quantum SWAP gates. The parity-check QNDs and SWAP gates in our hyper-EPP are constructed in an interesting structure, which can transfer the infidelity originating from the imperfect single-photon scattering into a heralded failure. Once the
parity-check QNDs and SWAP gates succeed, their fidelities will approach unity, which makes our hyper-EPP works faithfully. Moreover, our hyper-EPP is suitable to purify the photon systems entangled in a single DOF, such as the polarization DOF or the spatial-mode DOF by utilizing our fidelity-robust parity-check QND for two-photon systems in the polarization DOF or the spatial-mode DOF, and meanwhile it can be directly extended to purify the photon systems hyperentangled in more DOFs, such as the polarization DOF and the multiple-spatial-mode DOFs by constructing the corresponding parity-check QNDs and SWAP gates in a similar way. In principle, the fidelity of photon systems can gradually approach unity in all DOFs by performing multiple-round purification with our hyper-EPP.

\begin{figure}[th]
\centering
\includegraphics[width=7.2cm,angle=0]{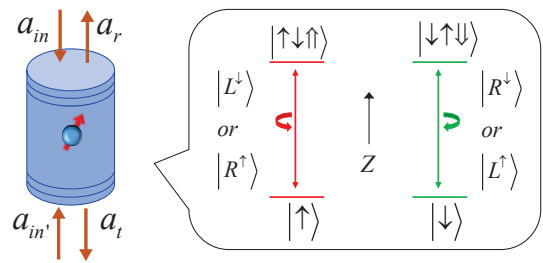}
\caption{Schematic diagrams for a singly charged QD in a
double-sided microcavity, the relevant energy levels, and the
optical transitions.
$|R^{\uparrow}\rangle$ ($|R^{\downarrow}\rangle$) and
$|L^{\uparrow}\rangle$ ($|L^{\downarrow}\rangle$) represent the right- and left-circularly polarized photons
propagating along (against) the quantization axis $Z$,
respectively.} \label{fig1}
\end{figure}

\section{Interaction rules for a circularly polarized photon
with a circuit unit consisting of a QD in a double-sided microcavity}  \label{sec2}

A singly charged quantum dot (QD) embedded in a double-sided microcavity constitutes an
interesting interface for a single QD and a single photon \cite{QD001,QD002}.
The microcavity consists of two partially reflected Bragg mirrors and supports
two circularly polarized degenerate modes. The
dipole transition associated with the singly charged QD is
strictly constrained by Pauli's exclusion principle, shown in Fig.~\ref{fig1}.  Here, the two single-electron ground states
$|\uparrow\rangle$ and $|\downarrow\rangle$ couple to the excited trion $X^{-}$ states $|\uparrow\downarrow\Uparrow\rangle$ and
$|\downarrow\uparrow\Downarrow\rangle$ by the absorption of a circularly polarized photon of $S_{z}=+1$
($|L^{\downarrow}\rangle$ or $|R^{\uparrow}\rangle$) and $S_{z}=-1$ ( $|R^{\downarrow}\rangle$ or $|L^{\uparrow}\rangle$), respectively.
The quantization axis \emph{Z} of the QD electron spin is along the direction of the microcavity.

When a circularly polarized photon is injected into
the double-sided microcavity, it will be scattered by the QD-cavity union.
The corresponding reflection and transmission coefficients could be
determined by solving the Heisenberg equations for
the cavity mode $\hat{a}$ and the QD operator $\hat{\sigma}_-$
in the interaction picture, along with the
input-output formalism of quantum optics \cite{Walls}
\begin{eqnarray}\label{eq1}
\begin{split}
\dot{\hat{a}}=&\;-[i(\omega_{c}-\omega)+\kappa+\frac{\kappa_{s}}{2}]\hat{a}
-g\hat{\sigma}_{-}\\
&\;-\sqrt{\kappa}(\hat{a}_{in'}+\hat{a}_{in})+\hat{H},\\
\dot{\hat{\sigma}}_{-}=&\;-[i(\omega_{X^{-}}-{\omega})+\frac{\gamma}{2}]\hat{\sigma}_{-}
-g\hat{\sigma}_{z}\hat{a}+\hat{G},\\
\hat{a}_{r}=&\;\hat{a}_{in}+\sqrt{\kappa}\hat{a},\\
\hat{a}_{t}=&\;\hat{a}_{in'}+\sqrt{\kappa}\hat{a},
\end{split}
\end{eqnarray}
where $\omega$, $\omega_{c}$, and $\omega_{X^{-}}$ represent
frequencies of the photon, the cavity, and $X^{-}$ transition,
respectively. $\gamma$, $\kappa$, and $\kappa_{s}$ are the $X^{-}$
dipole decay rate, the cavity decay rate, and the cavity side-leakage rate,
respectively. $\hat{S}$ and $\hat{G}$ are the noise operators  that  help  to  preserve  the
desired commutation relations. $\hat{a}_{in}$, $\hat{a}_{in'}$, $\hat{a}_{r}$, and
$\hat{a}_{t}$ are, respectively, the input and output modes, shown in Fig.~\ref{fig1}.

In the weak excitation limit where the QD
dominantly occupies the ground state
and $\langle\hat{\sigma}_{z}\rangle\simeq-1$, the reflection and transmission coefficients $r(\omega)$ and $t(\omega)$ of
the QD-cavity system can be detailed as \cite{Hu1}
\begin{eqnarray}\label{eq2}
\begin{split}
&r(\omega)=1+t(\omega),\\
&t(\omega)=-\frac{-\kappa[i(\omega_{X^{-}}-\omega)+\frac{\gamma}{2}]}{[i(\omega_{X^{-}}-\omega)
+\frac{\gamma}{2}][i(\omega_{c}-\omega)+\kappa+\frac{\kappa_{s}}{2}]+g^{2}}.
\end{split}
\end{eqnarray}

When a polarized photon is scattered by a QD initialized to the orthogonal ground state,
 it will decouple from the QD transition and
$g=0$, the reflection and transmission coefficients now can be simplified to
\begin{eqnarray}\label{eq3}
\begin{split}
&r_{0}(\omega)=1+t_{0}(\omega),\\
&t_{0}(\omega)=-\frac{\kappa}{i(\omega_{c}-\omega)+\kappa+\frac{\kappa_{s}}{2}}.
\end{split}
\end{eqnarray}
Therefore, when a circularly
polarized photon is scattered by a singly charged QD embedded in a double-sided microcavity,
the corresponding evolutions can be detailed as follows  \cite{Hu1}:
\begin{eqnarray}\label{eq4}
\begin{split}
&|R^{\uparrow}\uparrow\rangle\rightarrow r|L^{\downarrow}\uparrow\rangle
+t|R^{\uparrow}\uparrow\rangle,\\
&|L^{\downarrow}\uparrow\rangle\rightarrow r|R^{\uparrow}\uparrow\rangle
+t|L^{\downarrow}\uparrow\rangle,\\
&|R^{\downarrow}\uparrow\rangle\rightarrow t_{0}|R^{\downarrow}\uparrow\rangle
+r_{0}|L^{\uparrow}\uparrow\rangle,\\
&|L^{\uparrow}\uparrow\rangle\rightarrow t_{0}|L^{\uparrow}\downarrow\rangle
+r_{0}|R^{\downarrow}\uparrow\rangle,\\
&|R^{\downarrow}\downarrow\rangle\rightarrow r|L^{\uparrow}\downarrow\rangle
+t|R^{\downarrow}\downarrow\rangle,\\
&|L^{\uparrow}\downarrow\rangle\rightarrow r|R^{\downarrow}\downarrow\rangle
+t|L^{\uparrow}\downarrow\rangle,\\
&|R^{\uparrow}\downarrow\rangle\rightarrow t_{0}|R^{\uparrow}\downarrow\rangle
+r_{0}|L^{\downarrow}\downarrow\rangle,\\
&|L^{\downarrow}\downarrow\rangle\rightarrow t_{0}|L^{\downarrow}\downarrow\rangle
+r_{0}|R^{\uparrow}\downarrow\rangle.
\end{split}
\end{eqnarray}

\begin{figure}[!th]
\centering
\includegraphics[width=8cm,angle=0]{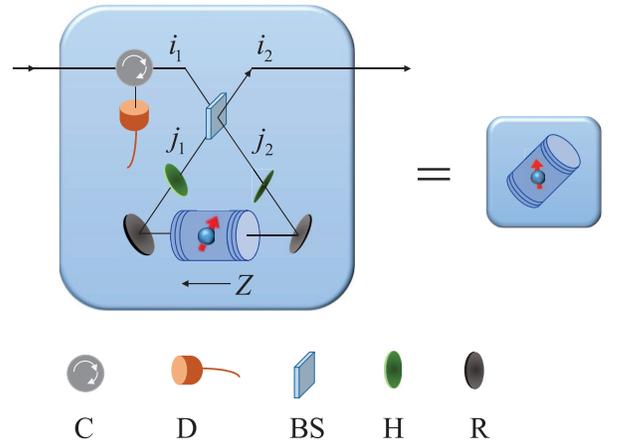}
\caption{ Schematic diagram of the modified QD-cavity union. BS is a $50:50$ beam
splitter, which performs the Hadamard operation,
$|i_{1}\rangle\rightarrow{(|j_{1}\rangle+|j_{2}\rangle)}/\sqrt{2}$ and
$|i_{2}\rangle\rightarrow(|j_{1}\rangle-|j_{2}\rangle)/\sqrt{2}$,
on the spatial-mode DOF of single photons. H represents a linear optical element, which performs the Hadamard operation,
$|R\rangle\rightarrow{(|R\rangle+|L\rangle)}/\sqrt{2}$ and
$|L\rangle\rightarrow(|R\rangle-|L\rangle)/\sqrt{2}$,
on the polarization DOF of single photons. R is a mirror, D is a
single-photon detector, and C is an optical circulator which can
route the photon into an appropriate path.} \label{fig2}
\end{figure}

By introducing several linear optical elements, we can modify the above QD-cavity union into a
robust QD-cavity unit, shown in Fig.~\ref{fig2}. As considered in previous works~\cite{BSA1,BSA2,BSA3,BSA4}, all linear optical elements are supposed to work  efficiently and complete ideal unitary transformations.
The electron spin of the QD is initially prepared in the state
$|\varphi^{+}\rangle=\frac{1}{\sqrt{2}}(|\uparrow\rangle+|\downarrow\rangle)$
and a right-circularly polarized photon $|R\rangle$ is injected into
the circuit unit. After the photon is performed on Hadamard operations in both the spatial-mode DOF and the polarization DOF, respectively, by a beam splitter (BS) and a linear optical element (H), it will be reflected by either mirror (R), and then impinge on the QD-cavity
system from the input ports. After scattering by the QD-cavity system, the photon with different polarizations will be reflected by either mirror {R}, pass through {H}, and then combine again at the BS. One can easily find that the photon output into mode $i_{2}$ of the BS
is the left-circularly polarized $|L\rangle$ and the corresponding probability amplitude is $T=\frac{1}{2}(t+r-t_{0}-r_{0})$, while the one into mode $i_1$ is the right-circularly polarized $|R\rangle$ and the corresponding probability amplitude is $D=\frac{1}{2}(t+r+t_{0}+r_{0})$.
For the photon in mode $i_1$ , it will pass through a circulator {C} and click the single-photon detector {D}. Therefore, the state of the hybrid system consisting of the photon and the QD
before the click of the detector {D} can be described as
\begin{eqnarray}\label{eq5}
\begin{split}
|\Phi\rangle_{1}=D|R\rangle|i_{1}\rangle|\varphi^{+}\rangle
+T|L\rangle|i_{2}\rangle|\varphi^{-}\rangle,
\end{split}
\end{eqnarray}
where the electron spin states $|\varphi^{\pm}\rangle=\frac{1}{\sqrt{2}}(|\uparrow\rangle\pm|\downarrow\rangle)$.

If the electron spin of the QD is initialized to the state $|\varphi^{-}\rangle$,
and the input photon is still the right-circularly polarized $|R\rangle$, the hybrid system consisting of the photon and the QD will evolve in a similar way as that detailed above.
The final state of the hybrid system before the photon is detected will be described as
\begin{eqnarray}\label{eq6}
\begin{split}
|\Phi\rangle_{2}=D|R\rangle|i_{1}\rangle|\varphi^{-}\rangle
+T|L\rangle|i_{2}\rangle|\varphi^{+}\rangle.
\end{split}
\end{eqnarray}

That is, our robust QD-cavity unit is identical to a modified QD-cavity union,
shown in Fig.~\ref{fig2}, which scatters a  right-circularly polarized photon $|R\rangle$
into two output modes no matter what the QD state is. If the scattered photon is in mode $i_{1}$,
it will click the single-photon detector and the electron spin of the QD in the double-sided microcavity will remain unchanged for a direct restarting round; while the photon
scattered  into mode $i_2$ evolves to a left-circularly
polarized photon $|L\rangle$ and completes a phase-flip operation on
the QD-confined electron spin, shown in Eqs.~(\ref{eq5}) and (\ref{eq6}). Based on
the new single-photon single-QD quantum interface, we propose three
fidelity-robust quantum circuits and they could be used directly in the
failure-heralded quantum computing and quantum networks.

\begin{figure}[!ht]
\begin{center}
\includegraphics[width=7.0cm,angle=0]{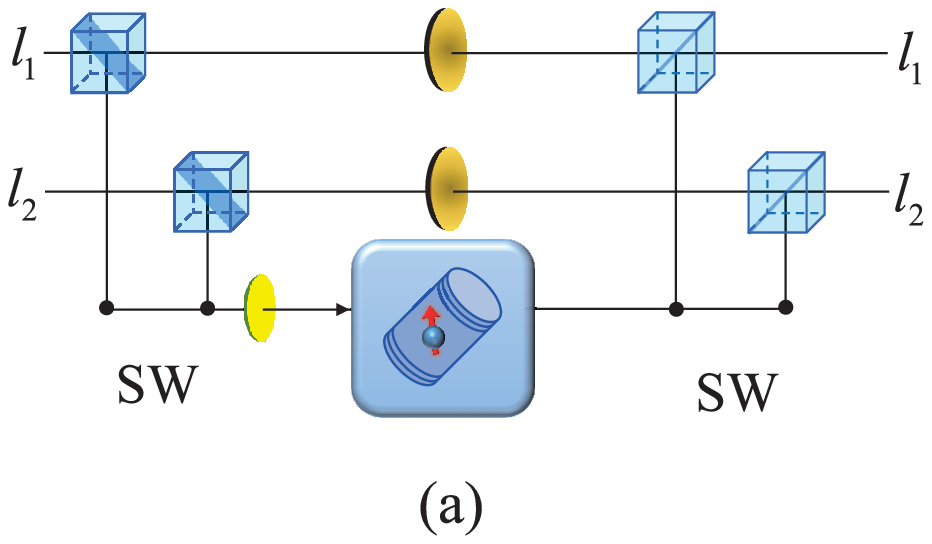} \hspace{30pt}
\includegraphics[width=7.0cm,angle=0]{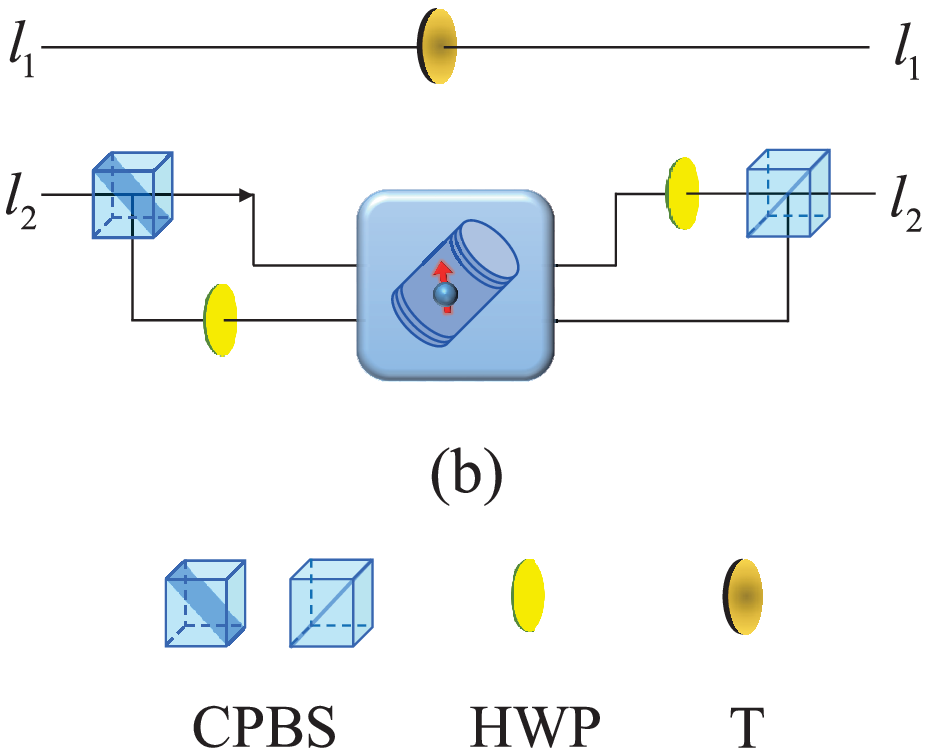}
\caption{(a) Schematic diagram of the fidelity-robust parity-check QND for a
two-photon system in the polarization DOF. (b) Schematic diagram of
the fidelity-robust parity-check QND for a two-photon system in the spatial-mode DOF.
CPBS is a circularly polarizing beam splitter, which transmits the
photon in the right-circular polarization $|R\rangle$ and reflects
the photon in the left-circular polarization $|L\rangle$,
respectively. HWP is a half-wave plate, which performs a bit-flip
operation $\sigma_{x}=|R\rangle\langle L|+|L\rangle\langle R|$ on
the photon in the polarization DOF. T is a partially transmitted mirror
with the transmission coefficient $T$. SW is an optical switch, which
makes the wave packets in different spatial modes
impinging into and exiting out of the circuit unit in sequence.}
\label{fig3}
\end{center}
\end{figure}

\section{Fidelity-robust parity-check QND for photon systems} \label{sec3}

A hyperentangled Bell state of a two-photon system in the
polarization and spatial-mode DOFs can be written as
\begin{equation}\label{eq7}
|\Psi\rangle=|\psi\rangle^{P}\otimes|\psi\rangle^{S},
\end{equation}
where $|\psi\rangle^{P}$ is one of the four Bell states in the polarization DOF,
\begin{eqnarray}\label{eq8}
\begin{split}
&|\phi^{\pm}\rangle^{P}=\frac{1}{\sqrt{2}}(|RR\rangle\pm|LL\rangle),\\
&|\psi^{\pm}\rangle^{P}=\frac{1}{\sqrt{2}}(|RL\rangle\pm|LR\rangle),
\end{split}
\end{eqnarray}
$|\psi\rangle^{S}$ is one of the following
four Bell states in the spatial-mode DOF,
\begin{eqnarray}\label{eq9}
\begin{split}
&|\phi^{\pm}\rangle^{S}=\frac{1}{\sqrt{2}}(|a_{1}c_{1}\rangle\pm|a_{2}c_{2}\rangle),\\
&|\psi^{\pm}\rangle^{S}=\frac{1}{\sqrt{2}}(|a_{1}c_{2}\rangle\pm|a_{2}c_{1}\rangle).
\end{split}
\end{eqnarray}
The parity-check QND for two hyper-encoded photons consists of two
 individual procedures. One  distinguishes the odd-parity states
$|\psi^{\pm}\rangle^{P}$ from the even-parity states
$|\phi^{\pm}\rangle^{P}$ in the polarization DOF and the other distinguishes the
odd-parity states $|\psi^{\pm}\rangle^{S}$ from the even-parity states
$|\phi^{\pm}\rangle^{S}$ in the spatial-mode DOF.

To measure the polarization parity of the two photons $A$ and $C$, one  initializes the
QD-confined electron, shown in Fig.~\ref{fig3}(a), to the state
$|\varphi^{+}\rangle=\frac{1}{\sqrt{2}}(|\uparrow\rangle+|\downarrow\rangle)$,
and then subsequently inputs photons $A$ and $C$ into the quantum
circuit. As shown in Fig.~\ref{fig3}(a),  the
left-circularly polarization component $|L\rangle$ in both the spatial modes $l_1$
and $l_2$ ($l=a$, $c$) will pass through the half-wave plate (HWP) that performs
a bit-flip operation on the photon, and then be scattered by the
modified QD-cavity union with the reflection coefficient $D$ and the transmission
coefficient $T$, detailed in Eqs.~(\ref{eq5}) and (\ref{eq6}); while
 the right-circularly polarization component $|R\rangle$
in both spatial modes will pass through the partially transmitted
mirror with the transmission coefficient $T$.

During the two single-photon scattering processes, if the photon $A$ ($C$) is reflected by the modified QD-cavity union, it will trigger the single-photon detector D, which marks the failure of the parity-check QND. If both photons $A$ and $C$ are transmitted through the modified QD-cavity union, there is no click of the single-photon detector D. Then, the two components of each photon, passing though different optical paths,
will combine again at a CPBS and be output into the spatial modes $l_1$ and $l_2$,
completing the QND on the polarization parity.
The evolutions of the whole system consisting of photons $AC$ and
the electron are described as follows:
\begin{eqnarray}\label{eq10}
\begin{split}
&|\phi^{\pm}\rangle^{P}|\psi\rangle^{S}\otimes|\varphi^{+}\rangle\rightarrow T^{2}
|\phi^{\pm}\rangle^{P}|\psi\rangle^{S}|\varphi^{+}\rangle,\\
&|\psi^{\pm}\rangle^{P}|\psi\rangle^{S}\otimes|\varphi^{+}\rangle\rightarrow T^{2}
|\psi^{\pm}\rangle^{P}|\psi\rangle^{S}|\varphi^{-}\rangle.
\end{split}
\end{eqnarray}
where
$|\varphi^{-}\rangle=\frac{1}{\sqrt{2}}(|\uparrow\rangle-|\downarrow\rangle)$
represents a phase-flip state compared to $|\varphi^{+}\rangle$.
By measuring the state of the electron in the
orthogonal basis $\{|\varphi^{+}\rangle,|\varphi^{-}\rangle\}$, the
polarization parity of the two photons can be
determined. In detail, if the electron is in
state $|\varphi^{+}\rangle$, the polarization parity of photons $AC$ will be projected into the
even subspace spanned by $|\phi^{\pm}\rangle^{P}$; if
the electron is in  state $|\varphi^{-}\rangle$, the polarization parity of $AC$ will be  projected into the
odd subspace spanned by $|\psi^{\pm}\rangle^{P}$.

To measure the spatial-mode parity of the two photons $A$ and $C$, one  initializes the QD-confined
electron, shown in Fig.~\ref{fig3}(b), to the state
$|\varphi^{+}\rangle$, and then subsequently inputs photons $A$ and $C$ into the quantum
circuit. As shown in Fig.~\ref{fig3}(b), the left-circularly polarization component $|L\rangle$ in
spatial mode $l_{2}$ will subsequently pass through the HWP
and  be scattered by the modified QD-cavity union, while the right-circularly polarization component $|R\rangle$ in
spatial mode $l_{2}$ will first be scattered by the modified QD-cavity union and then pass through HWP. If both photons are transmitted through the modified QD-cavity union, there is no click of the single-photon detector D, and the two components discussed above will combine at a CPBS and then be output into the spatial mode $l_{2}$.
However, for the spatial mode $l_{1}$, both the left- and right-circularly polarized components will pass through a partially transmitted mirror with the transmission coefficient $T$.

The evolutions of the whole system consisting of the photons $AC$ and
the electron are described as
\begin{eqnarray}\label{eq11}
\begin{split}
&|\psi\rangle^{P}|\phi^{\pm}\rangle^{S}\otimes|\varphi^{+}\rangle\rightarrow T^{2}
|\psi\rangle^{P}|\phi^{\pm}\rangle^{S}\otimes|\varphi^{+}\rangle,\\
&|\psi\rangle^{P}|\psi^{\pm}\rangle^{S}\otimes|\varphi^{+}\rangle\rightarrow T^{2}
|\psi\rangle^{P}|\psi^{\pm}\rangle^{S}\otimes|\varphi^{-}\rangle.
\end{split}
\end{eqnarray}
To complete the spatial-mode parity check
on photons $AC$, one measures the electron in the
orthogonal basis $\{|\varphi^{+}\rangle,|\varphi^{-}\rangle\}$.
If the electron is in state $|\varphi^{+}\rangle$, the photons $AC$ are projected into the  even-spatial-mode-parity subspace spanned by $|\phi^{\pm}\rangle^{S}$; if the electron is in state $|\varphi^{-}\rangle$, the photons $AC$ are projected into the odd-spatial-mode-parity subspace spanned by $|\psi^{\pm}\rangle^{S}$.

The failure of the parity-check QNDs for photons, hyper encoded in the polarization and spatial-mode DOFs, can be heralded by the clicks of the single-photon detectors in the modified QD-cavity unions. However, once the parity-check QNDs succeed, their fidelities approach unity even
when  nonideal scattering  conditions are used.
The original infidelity term is transformed into the
failure heralded by the click of single-photon detectors. The error-free component in each process of the parity-check QNDs is proportional to the square of the transmission coefficient $T$. The efficiency of each parity-check QND for photons in the polarization or spatial-mode DOF is $\eta_{PC}=|T|^{4}$, which in principle approaches unity when the single-photon scattering is ideal.

\begin{figure}[!ht]
\begin{center}
\includegraphics[width=8.0cm,angle=0]{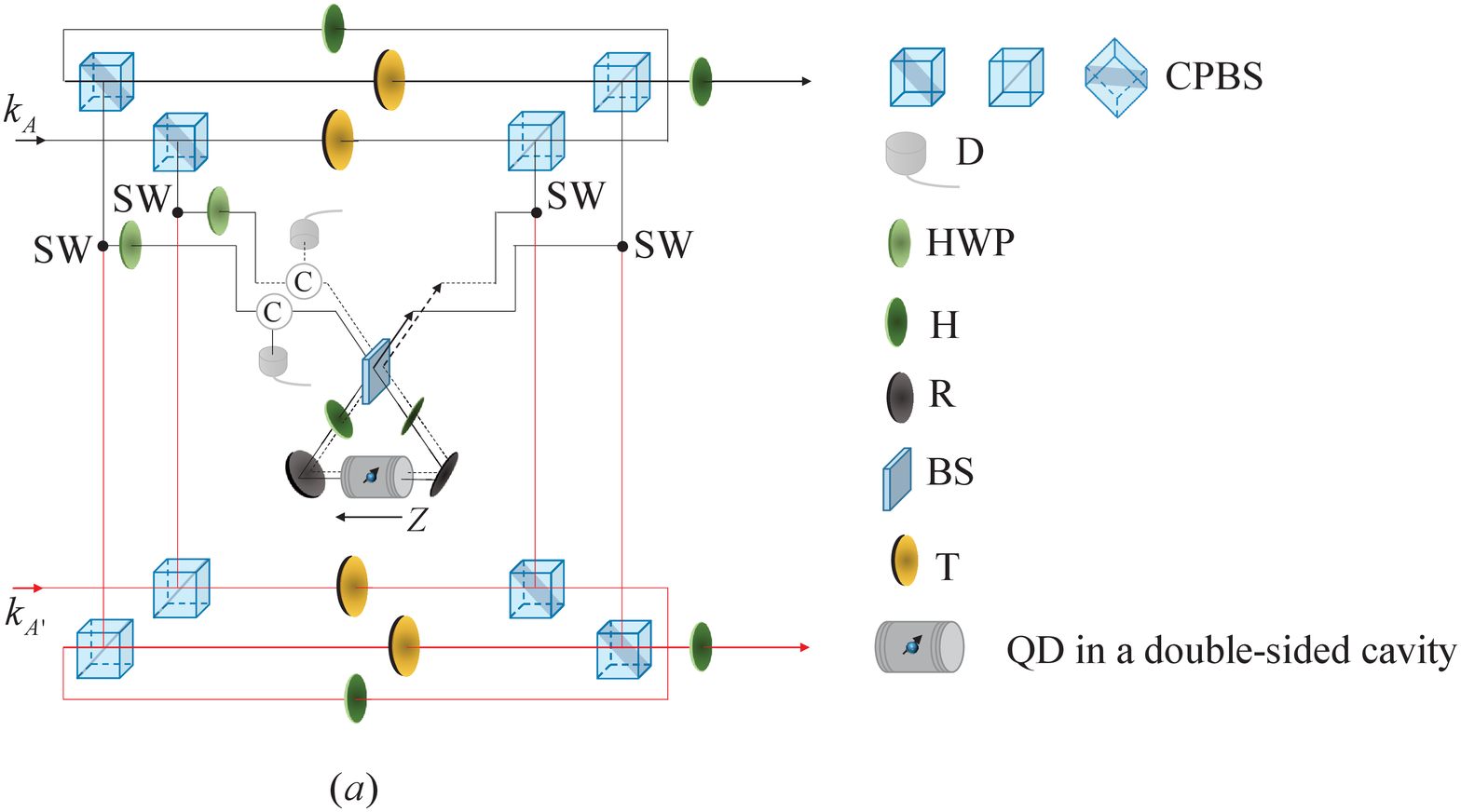} \hspace{30pt}
\includegraphics[width=6.6cm,angle=0]{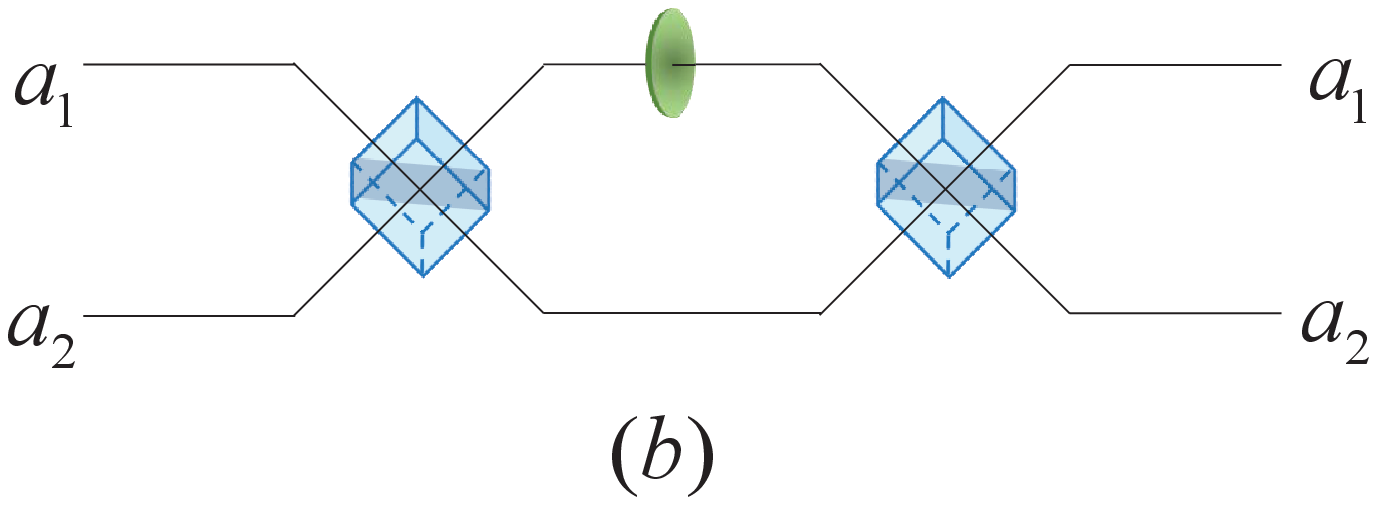}
\caption{(a) Schematic diagram of the fidelity-robust SWAP gate between the
polarization states of the photons $A$ and $A'$ (P-P SWAP gate). $k~(k=a_{1},
a_{2})$ represents either spatial mode of the photons $A$ and $A'$. (b) Schematic
diagram of the SWAP gate between the spatial-mode state and polarization
state of the same photon (P-S SWAP gate).} \label{fig4}
\end{center}
\end{figure}

\section{Fidelity-robust SWAP gates} \label{sec4}

\subsection{Fidelity-robust SWAP gate between polarization states of two photons}

Suppose two photons $A$ and $A'$ hyper encoded in the polarization and spatial-mode DOFs are
\begin{eqnarray}\label{eq12}
\begin{split}
&|\Phi\rangle_{A}=(\alpha_{1}|R\rangle+\beta_{1}|L\rangle)_{A}\otimes|\Phi_{k}\rangle_{A},\\
&|\Phi\rangle_{A'}=(\alpha_{2}|R\rangle+\beta_{2}|L\rangle)_{A'}\otimes|\Phi_{k}\rangle_{A'},
\end{split}
\end{eqnarray}
where
$|\Phi_{k}\rangle_{A}=(\gamma_{1}|a_{1}\rangle+\delta_{1}|a_{2}\rangle)_{A}$
and
$|\Phi_{k}\rangle_{A'}=(\gamma_{2}|a_{1}\rangle+\delta_{2}|a_{2}\rangle)_{A'}$
are the spatial-mode states of the photons $A$ and $A'$. A perfect SWAP gate (P-P
SWAP gate) between the photons $AA'$ will exchange the polarization states
of the photons $A$ and $A'$ without affecting their spatial-mode states.
The ideal output of the P-P SWAP gate will be
\begin{eqnarray}\label{eq15-i}
|\Phi\rangle=&&\!\!\!(\alpha_{2}|R\rangle+\beta_{2}|L\rangle)_{A}|\Phi_{k}\rangle_{A}
\otimes(\alpha_{1}|R\rangle\nonumber\\
&+&\beta_{1}|L\rangle)_{A'}|\Phi_{k}\rangle_{A'}.
\end{eqnarray}

The schematic diagram of our  P-P
SWAP gate is shown in Fig.~\ref{fig4}(a). Here, we use $k_{A}$ and $k_{A'}$ to
represent the spatial modes of the  photons ${A}$ and ${A'}$, since both the spatial modes $a_{1}$ and $a_{2}$ of
the photon $A$ ($A'$) pass through the same optical path.

To implement the  P-P SWAP gate,
one initializes the QD-confined electron to the state
$|\varphi^{+}\rangle$, and then inputs the photons $A$ and $A'$ through $k_{A}$ and
$k_{A'}$ into the quantum circuit, respectively. For the first round, the CPBS splits the photon into two components of orthogonal circular polarizations.
The left-circularly polarized one will  subsequently pass through the HWP and be scattered by the  modified QD-cavity
union, while the right-circularly polarized one will pass through the partially transmitted mirror with the transmission
coefficient $T$. If the single-photon detector clicks, the P-P SWAP gate fails.
If both photons are transmitted through the modified QD-cavity union, the P-P SWAP gate process will continue and combine the
two components in left- and right-circular
polarization at another CPBS. After that, one performs Hadamard operations on the
polarization DOF of both the photons $AA'$ and the QD-confined
electron, respectively. The state of the system consisting of the
two photons and the electron is transformed into
\begin{eqnarray}\label{eq13}
\begin{split}
|\Phi\rangle_{1}=&\frac{T^{2}}{2}[\alpha_{1}\alpha_{2}(|RR\rangle+|RL\rangle
+|LR\rangle+|LL\rangle)|\uparrow\rangle\\
&+\alpha_{1}\beta_{2}(|RR\rangle-|RL\rangle+|LR\rangle-|LL\rangle)|\downarrow\rangle\\
&+\beta_{1}\alpha_{2}(|RR\rangle+|RL\rangle-|LR\rangle-|LL\rangle)|\downarrow\rangle\\
&+\beta_{1}\beta_{2}(|RR\rangle-|RL\rangle-|LR\rangle+|LL\rangle)|\uparrow\rangle]\\
&\otimes|\Phi_{k}\rangle_{A}\otimes|\Phi_{k}\rangle_{A'}.
\end{split}
\end{eqnarray}

Subsequently, the photons will be split again for the second round, and
each component passes through a similar optical path as it does in the previous procedure.
For the left-circularly polarized one, it will first pass through the HWP and then be scattered by the
modified QD-cavity union for the second time.
For the right-circularly polarized one, it will
pass through the partially transmitted mirror with the
transmission coefficient $T$. If the photons can still be transmitted through the modified QD-cavity union, the process is
going on and the two wave packets in the left- and right-circular
polarizations reunion again at another CPBS. After Hadamard operations performed on the
photons $AA'$ and the electron, the state of the whole system is transformed into
\begin{eqnarray}\label{eq14}
\begin{split}
|\Phi\rangle_{2}=&\frac{T^{4}}{\sqrt{2}}[(\alpha_{2}|R\rangle
+\beta_{2}|L\rangle)_{A}(\alpha_{1}|R\rangle+\beta_{1}|L\rangle)_{A'}|\uparrow\rangle\\
&+(\alpha_{2}|R\rangle-\beta_{2}|L\rangle)_{A}(\alpha_{1}|R\rangle
-\beta_{1}|L\rangle)_{A'}|\downarrow\rangle]\\
&\otimes|\Phi_{k}\rangle_{A}\otimes|\Phi_{k}\rangle_{A'}.
\end{split}
\end{eqnarray}

Finally, to complete the P-P SWAP gate, one  measures the electron
in the basis $\{|\uparrow\rangle,|\downarrow\rangle\}$ and exchanges the polarization states
of the photons $A$  and $A'$ perfectly as shown in Eq.~(\ref{eq15-i}) up to a single-qubit feedback,
conditioned on the outcomes of the electron measurement.
If the outcome of the electron measurement is $|\uparrow\rangle$, the state of the two photons is
transformed into
\begin{eqnarray}\label{eq15}
\begin{split}
|\Phi\rangle_{f1}=T^{4}(\alpha_{2}|R\rangle+\beta_{2}|L\rangle)_{A}|\Phi_{k}\rangle_{A}\\
\otimes(\alpha_{1}|R\rangle+\beta_{1}|L\rangle)_{A'}|\Phi_{k}\rangle_{A'}.
\end{split}
\end{eqnarray}
If the outcome of the electron measurement is  $|\downarrow\rangle$,
one needs to perform a phase-flip operation $\sigma_{z}=|R\rangle\langle
R|-|L\rangle\langle L|$  on both the photons $A$ and $A'$ to evolve the
photons $AA'$ into the above state shown in  Eq.~(\ref{eq15}).

In our P-P SWAP gate, the click of the single-photon
detector alarms the failure of the gate.
Once the P-P SWAP gate succeeds, its fidelity approaches unity as the parity-check QND does, since the practical reflection and transition coefficients appear as a global coefficient in the
final state $|\Phi\rangle_{f1}$, which means the success  efficiency
of the P-P SWAP gate is $\eta_{SWAP}=|T|^{8}$.

\subsection{Hybrid SWAP gate between the spatial-mode and polarization states of the same photon}

The schematic diagram for implementing the hybrid SWAP gate between
the  polarization  and spatial-mode states (P-S SWAP) of the same photon is shown in
Fig.~\ref{fig4}(b). Suppose photon $A$ is in state
$|\Phi\rangle_{A}=(\alpha_{1}|R\rangle+\beta_{1}|L\rangle)(\gamma_{1}|a_{1}\rangle
+\delta_{1}|a_{2}\rangle)$. One injects the spatial modes $a_{1}$ and $a_{2}$ of photon $A$
into the corresponding input ports $a_{1}$ and $a_{2}$ of the quantum circuit, shown in
Fig.~\ref{fig4}(b).
The state of the photon $A$ will evolve into
\begin{eqnarray}\label{eq16}
\begin{split}
|\Phi\rangle_{f2}=(\alpha_{1}|a_{1}\rangle+\beta_{1}|a_{2}\rangle)
(\gamma_{1}|R\rangle+\delta_{1}|L\rangle),
\end{split}
\end{eqnarray}
which is the ideal target state of the hybrid P-S SWAP gate, and the corresponding efficiency equals unity.

\section{Faithful EPP for two-photon systems hyperentangled in two DOFs} \label{sec5}

In the practical transmission of photons in hyperentangled
Bell states for high-capacity quantum communication, both the
bit-flip and phase-flip errors will occur randomly. Usually,
a phase-flip error can be transformed into a bit-flip error using a bilateral local operation. The bit-flip error purification can, therefore, be used to decrease both the bit-flip and phase-flip errors. Below, we only  discuss the purification for bit-flip errors of two-photon hyperentangled mixed states.

The two identical photon pairs $AB$ and $CD$ in mixed hyperentangled
Bell states  with the bit-flip errors in both the polarization and spatial-mode DOFs can be described as
\begin{equation}\label{eq17}
\begin{split}
\rho_{AB}\;=\;&[F_{1}|\phi^{+}\rangle_{AB}^{P} \langle \phi^{+}|
+(1-F_{1})|\psi^{+}\rangle_{AB}^{P} \langle \psi^{+}|]\\
&\otimes[F_{2}|\phi^{+}\rangle_{AB}^{S} \langle \phi^{+}|
+(1-F_{2})|\psi^{+}\rangle_{AB}^{S} \langle \psi^{+}|], \\
\rho_{CD}\;=\;&[F_{1}|\phi^{+}\rangle_{CD}^{P} \langle \phi^{+}|
+(1-F_{1})|\psi^{+}\rangle_{CD}^{P} \langle \psi^{+}|]\\
&\otimes [F_{2}|\phi^{+}\rangle_{CD}^{S} \langle \phi^{+}|
+(1-F_{2})|\psi^{+}\rangle_{CD}^{S} \langle \psi^{+}|].
\end{split}
\end{equation}
Here, the subscripts $A$ and $C$ denote the photons sent to  Alice and
the subscripts $B$ and $D$ denote the photons sent to Bob.
We use the fidelity $F=F_{1}F_{2}$ to characterize the mixed hyperentangled Bell states, where $F_{1}$ and $F_{2}$ represent the corresponding probabilities of the states
$|\phi^{+}\rangle_{AB}^{P}$  $(|\phi^{+}\rangle_{CD}^{P})$ and
$|\phi^{+}\rangle_{AB}^{S}$  $ (|\phi^{+}\rangle_{CD}^{S})$ without bit-flip errors in the mixed hyperentangled Bell states, respectively. Hyperentanglement witness is another measurement to characterize hyperentanglement \cite{chara1}, and the nonlocal character of a hyperentangled state can also be verified by testing every DOF against the Bell-Clauser-Horne-Shimony-Holt inequality \cite{hyper4}.

\begin{figure}[!th]
\centering
\includegraphics[width=8cm,angle=0]{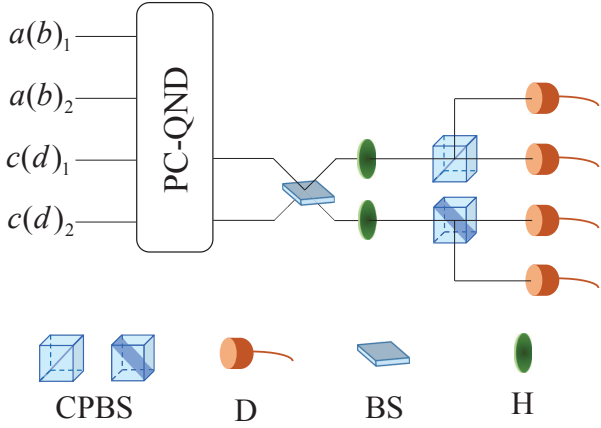}
\caption{Schematic diagram of the first purification step of the hyper-EPP.
BS is used to perform the Hadamard operation
$[|c(d)_{1}\rangle\rightarrow\frac{1}{\sqrt{2}}(|c(d)_{1}\rangle
+|c(d)_{2}\rangle),|c(d)_{2}\rangle\rightarrow\frac{1}{\sqrt{2}}(|c(d)_{1}\rangle
-|c(d)_{2}\rangle)]$ on single photons
in the spatial-mode DOF.} \label{fig5}
\end{figure}

Initially, the state of the system composed of the two identical
photon pairs $AB$ and $CD$ can be expressed as
$\rho_{0}=\rho_{AB}\otimes\rho_{CD}$, which is a
mixture of maximally hyperentangled pure states. In the polarization
DOF, it is a mixture of the states
$|\phi^{+}\rangle_{AB}^{P}\otimes|\phi^{+}\rangle_{CD}^{P}$,
$|\phi^{+}\rangle_{AB}^{P}\otimes|\psi^{+}\rangle_{CD}^{P}$,
$|\psi^{+}\rangle_{AB}^{P}\otimes|\phi^{+}\rangle_{CD}^{P}$, and
$|\psi^{+}\rangle_{AB}^{P}\otimes|\psi^{+}\rangle_{CD}^{P}$ with the
probabilities $F_{1}^{2}$, $F_{1}(1-F_{1})$, $(1-F_{1})F_{1}$, and
$(1-F_{1})^{2}$, respectively. In the spatial-mode DOF, it is a mixture
of the states
$|\phi^{+}\rangle_{AB}^{S}\otimes|\phi^{+}\rangle_{CD}^{S}$,
$|\phi^{+}\rangle_{AB}^{S}\otimes|\psi^{+}\rangle_{CD}^{S}$,
$|\psi^{+}\rangle_{AB}^{S}\otimes|\phi^{+}\rangle_{CD}^{S}$, and
$|\psi^{+}\rangle_{AB}^{S}\otimes|\psi^{+}\rangle_{CD}^{S}$ with the
probabilities $F_{2}^{2}$, $F_{2}(1-F_{2})$, $(1-F_{2})F_{2}$, and
$(1-F_{2})^{2}$, respectively.

Our faithful hyper-EPP for two-photon hyperentangled systems with the bit-flip errors in both the
polarization and spatial-mode DOFs is
divided into two purification steps. The first purification step is
achieved with the fidelity-robust parity-check QND in the polarization and
spatial-mode DOFs, as described in Sec.~\ref{sec3}. The second step
is based on two SWAP gates, as
introduced in Sec.~\ref{sec4}. As all quantum operations, the parity-check QNDs, and the SWAP gates work
with a near-unity fidelity, the hyper-EPP here will be performed faithfully and work without the secondary pollution
introduced by the infidelity of each quantum operation.

\subsection{The first purification step of the hyper-EPP}

The process of the first purification step of our hyper-EPP is shown in
Fig.~\ref{fig5}. In Alice's node, it can be disassembled into the
parity-check QNDs on the photons $AC$ in the  polarization and spatial-mode
DOFs, Hadamard operations on photon $C$ in the polarization and
spatial-mode DOFs, and the detection on photon $C$. In Bob's node, it can be disassembled into the same operations by replacing the photons  $A$ and $C$ with $B$ and $D$, respectively.

After Alice and Bob,  respectively, perform the parity-check QNDs on the photons $AC$ and
$BD$ in both DOFs, the outcomes of the measurement are shown in Table.~\ref{table1}.

\begin{table}[th]
\centering
\caption{The results of the parity-check QNDs on the photons $AC$ and $BD$.}
\begin{tabular}{cccc}\hline\hline
Case      & DOF               & Parity mode  & Bell states     \\\hline
(1)       & P                 & same
& $|\phi^{+}\rangle_{AB}^{P} |\phi^{+}\rangle_{CD}^{P}$
or $|\psi^{+}\rangle_{AB}^{P} |\psi^{+}\rangle_{CD}^{P}$ \\
          & S                 & same
          & $|\phi^{+}\rangle_{AB}^{S} |\phi^{+}\rangle_{CD}^{S}$
          or $|\psi^{+}\rangle_{AB}^{S} |\psi^{+}\rangle_{CD}^{S}$ \\\cline{1-4}
 (2)      & p                 & different
 & $|\phi^{+}\rangle_{AB}^{P} |\psi^{+}\rangle_{CD}^{P}$
 or $|\psi^{+}\rangle_{AB}^{P} |\phi^{+}\rangle_{CD}^{P}$ \\
          & S                 & different
          & $|\phi^{+}\rangle_{AB}^{S} |\psi^{+}\rangle_{CD}^{S}$
          or $|\psi^{+}\rangle_{AB}^{S} |\phi^{+}\rangle_{CD}^{S}$ \\\cline{1-4}
(3)       & P                 & different
& $|\phi^{+}\rangle_{AB}^{P} |\psi^{+}\rangle_{CD}^{P}$
or $|\psi^{+}\rangle_{AB}^{P} |\phi^{+}\rangle_{CD}^{P}$ \\
          & S                 & same
          & $|\phi^{+}\rangle_{AB}^{S} |\phi^{+}\rangle_{CD}^{S}$
          or $|\psi^{+}\rangle_{AB}^{S} |\psi^{+}\rangle_{CD}^{S}$ \\\cline{1-4}
(4)       & P                 & same
& $|\phi^{+}\rangle_{AB}^{P} |\phi^{+}\rangle_{CD}^{P}$
or $|\psi^{+}\rangle_{AB}^{P} |\psi^{+}\rangle_{CD}^{P}$ \\
          & S                 & different
          & $|\phi^{+}\rangle_{AB}^{S} |\psi^{+}\rangle_{CD}^{S}$
          or $|\psi^{+}\rangle_{AB}^{S} |\phi^{+}\rangle_{CD}^{S}$ \\ \hline\hline 
\end{tabular}\label{table1}
\end{table}

In case (1), the states of the photons $AC$ and $BD$ are of the same
parity in both the polarization and spatial-mode DOFs, which
corresponds to the state
$|\phi^{+}\rangle_{AB}^{P}|\phi^{+}\rangle_{CD}^{P}$ or
$|\psi^{+}\rangle_{AB}^{P}|\psi^{+}\rangle_{CD}^{P}$ in the polarization
DOF and the state
$|\phi^{+}\rangle_{AB}^{S}|\phi^{+}\rangle_{CD}^{S}$ or
$|\psi^{+}\rangle_{AB}^{S}|\psi^{+}\rangle_{CD}^{S}$ in the spatial-mode
DOF. In detail, if the parities of the photons $AC$ and $BD$ are even in both the polarization and spatial-mode DOFs, the states of the
four-photon system $ABCD$ can be expressed as
\begin{equation}\label{eq18}
\begin{split}
&|\Phi_{1}\rangle^{P}=\frac{1}{\sqrt{2}}(|RRRR\rangle+|LLLL\rangle)_{ABCD},\\
&|\Phi_{2}\rangle^{P}=\frac{1}{\sqrt{2}}(|RLRL\rangle+|LRLR\rangle)_{ABCD},\\
&|\Phi_{1}\rangle^{S}=\frac{1}{\sqrt{2}}(|a_{1}b_{1}c_{1}d_{1}\rangle
+|a_{2}b_{2}c_{2}d_{2}\rangle)_{ABCD},\\
&|\Phi_{2}\rangle^{S}=\frac{1}{\sqrt{2}}(|a_{1}b_{2}c_{1}d_{2}\rangle
+|a_{2}b_{1}c_{2}d_{1}\rangle)_{ABCD}.
\end{split}
\end{equation}
In contrast, if the parities of the photons $AC$ and $BD$ are odd
in both the polarization and spatial-mode DOFs, the states of the
four-photon system $ABCD$ can be expressed as
\begin{equation}\label{eq19}
\begin{split}
&|\Phi_{3}\rangle^{P}=\frac{1}{\sqrt{2}}(|RRLL\rangle+|LLRR\rangle)_{ABCD},\\
&|\Phi_{4}\rangle^{P}=\frac{1}{\sqrt{2}}(|RLLR\rangle+|LRRL\rangle)_{ABCD},\\
&|\Phi_{3}\rangle^{S}=\frac{1}{\sqrt{2}}(|a_{1}b_{1}c_{2}d_{2}\rangle
+|a_{2}b_{2}c_{1}d_{1}\rangle)_{ABCD},\\
&|\Phi_{4}\rangle^{S}=\frac{1}{\sqrt{2}}(|a_{1}b_{2}c_{2}d_{1}\rangle
+|a_{2}b_{1}c_{1}d_{2}\rangle)_{ABCD}.
\end{split}
\end{equation}
The states $|\Phi_{3}\rangle^{P}$ and $|\Phi_{4}\rangle^{P}$ can be
transformed into $|\Phi_{1}\rangle^{P}$ and $|\Phi_{2}\rangle^{P}$
by bit-flip operations on the photons $C$ and $D$ in the polarization DOF,
respectively. Similarly, the states $|\Phi_{3}\rangle^{S}$ and
$|\Phi_{4}\rangle^{S}$ can be transformed into
$|\Phi_{1}\rangle^{S}$ and $|\Phi_{2}\rangle^{S}$ by bit-flip
operations  on the photons $C$ and $D$ in the spatial-mode DOF, respectively.

Subsequently, Alice and Bob perform Hadamard operations on the
photons $C$ and $D$ in both the polarization and spatial-mode DOFs,
respectively. The states $|\Phi_{1}\rangle^{P}$,
$|\Phi_{2}\rangle^{P}$, $|\Phi_{1}\rangle^{S}$, and
$|\Phi_{2}\rangle^{S}$ are changed into the states
$|\Phi_{1}'\rangle^{P}$, $|\Phi_{2}'\rangle^{P}$,
$|\Phi_{1}'\rangle^{S}$, and $|\Phi_{2}'\rangle^{S}$, respectively.
Here,
\begin{equation}\label{eq19-1}
\begin{split}
|\Phi'_{1}\rangle^{P}=&\frac{1}{2\sqrt{2}}[(|RR\rangle+|LL\rangle)_{AB}
(|RR\rangle+|LL\rangle)_{CD}\\
&+(|RR\rangle-|LL\rangle)_{AB}(|RL\rangle+|LR\rangle)_{CD}],\\
|\Phi'_{2}\rangle^{P}=&\frac{1}{2\sqrt{2}}[(|RL\rangle+|LR\rangle)_{AB}
(|RR\rangle-|LL\rangle)_{CD}\\
&+(-|RL\rangle+|LR\rangle)_{AB}(|RL\rangle-|LR\rangle)_{CD}], \\
|\Phi'_{1}\rangle^{S}=&\frac{1}{2\sqrt{2}}[(|a_{1}b_{1}\rangle+|a_{2}b_{2}\rangle)_{AB}
(|c_{1}d_{1}\rangle+|c_{2}d_{2}\rangle)_{CD}\\
&+(|a_{1}b_{1}\rangle-|a_{2}b_{2}\rangle)_{AB}(|c_{1}d_{2}\rangle+|c_{2}d_{1}\rangle)_{CD}],\\
|\Phi'_{2}\rangle^{S}=&\frac{1}{2\sqrt{2}}[(|a_{1}b_{2}\rangle+|a_{2}b_{1}\rangle)_{AB}
(|c_{1}d_{1}\rangle-|c_{2}d_{2}\rangle)_{CD}\\
&+(-|a_{1}b_{2}\rangle+|a_{2}b_{1}\rangle)_{AB}(|c_{1}d_{2}\rangle-|c_{2}d_{1}\rangle)_{CD}].
\end{split}
\end{equation}

Finally, photons $C$ and $D$ are detected. If photons $C$
and $D$ are in even-parity state in the polarization DOF
(the spatial-mode DOF), the first purification step
of the hyper-EPP is finished directly. If photons $C$ and $D$ are in the
odd-parity state in the polarization DOF (the spatial-mode DOF), the
first step of the hyper-EPP is finished after performing a phase-flip
operation $\sigma_{z}^{P}=|R\rangle\langle R|-|L\rangle\langle
L|(\sigma_{z}^{S}=|b_{1}\rangle\langle b_{1}|-|b_{2}\rangle\langle
b_{2}|)$ on photon $B$ in the polarization DOF (the spatial-mode DOF).
In case (1), Alice and Bob will reserve the photon pair $AB$,
and it will contribute to the final purified state of the hyper-EPP.

In case (2), the states of the photons $AC$ and $BD$ are of
different parities in both the polarization and spatial-mode DOFs,
which corresponds to the state $|\phi^{+}\rangle_{AB}^{P}
|\psi^{+}\rangle_{CD}^{P}$ or $|\psi^{+}\rangle_{AB}^{P}
|\phi^{+}\rangle_{CD}^{P}$ in the polarization DOF and the state
$|\psi^{+}\rangle_{AB}^{S} |\phi^{+}\rangle_{CD}^{S}$ or
$|\phi^{+}\rangle_{AB}^{S} |\psi^{+}\rangle_{CD}^{S}$ in the
spatial-mode DOF. In detail, if the photons $AC$ are in the even-parity
states in both the polarization and spatial-mode DOFs and $BD$ are in
the odd-parity states in both the polarization and spatial-mode DOFs,
the states of the four-photon system $ABCD$ can be expressed as
\begin{equation}\label{eq20}
\begin{split}
&|\Phi_{5}\rangle^{P}=\frac{1}{\sqrt{2}}(|RRRL\rangle+|LLLR\rangle)_{ABCD},\\
&|\Phi_{6}\rangle^{P}=\frac{1}{\sqrt{2}}(|RLRR\rangle+|LRLL\rangle)_{ABCD},\\
&|\Phi_{5}\rangle^{S}=\frac{1}{\sqrt{2}}(|a_{1}b_{1}c_{1}d_{2}\rangle
+|a_{2}b_{2}c_{2}d_{1}\rangle)_{ABCD},\\
&|\Phi_{6}\rangle^{S}=\frac{1}{\sqrt{2}}(|a_{1}b_{2}c_{1}d_{1}\rangle
+|a_{2}b_{1}c_{2}d_{2}\rangle)_{ABCD}.
\end{split}\end{equation}
In contrast, if photons $AC$ are in the odd-parity states in both the
polarization and spatial-mode DOFs and $BD$ are in the even-parity
states in both the polarization and spatial-mode DOFs, the states of the
four-photon system $ABCD$ can be expressed as
\begin{equation}\label{eq21}
\begin{split}
&|\Phi_{7}\rangle^{P}=\frac{1}{\sqrt{2}}(|RRLR\rangle+|LLRL\rangle)_{ABCD},\\
&|\Phi_{8}\rangle^{P}=\frac{1}{\sqrt{2}}(|RLLL\rangle+|LRRR\rangle)_{ABCD},\\
&|\Phi_{7}\rangle^{S}=\frac{1}{\sqrt{2}}(|a_{1}b_{1}c_{2}d_{1}\rangle
+|a_{2}b_{2}c_{1}d_{2}\rangle)_{ABCD},\\
&|\Phi_{8}\rangle^{S}=\frac{1}{\sqrt{2}}(|a_{2}b_{1}c_{1}d_{1}\rangle
+|a_{1}b_{2}c_{2}d_{2}\rangle)_{ABCD}.
\end{split}\end{equation}
The states $|\Phi_{7}\rangle^{P}$ and $|\Phi_{8}\rangle^{P}$ can be
transformed into $|\Phi_{5}\rangle^{P}$ and $|\Phi_{6}\rangle^{P}$
by bit-flip operations  on the photons $C$ and $D$ in the polarization DOF,
respectively. Similarly, the states $|\Phi_{7}\rangle^{S}$ and
$|\Phi_{8}\rangle^{S}$ can be transformed into
$|\Phi_{5}\rangle^{S}$ and $|\Phi_{6}\rangle^{S}$ by bit-flip
operations on the photons $C$ and $D$ in the spatial-mode DOF, respectively.

Subsequently, after Alice and Bob, respectively, perform Hadamard operations on the
photons $C$ and $D$ in the polarization and spatial-mode DOFs,
the states $|\Phi_{5}\rangle^{P}$,
$|\Phi_{6}\rangle^{P}$, $|\Phi_{5}\rangle^{S}$, and
$|\Phi_{6}\rangle^{S}$ are changed into the states
$|\Phi_{5}'\rangle^{P}$, $|\Phi_{6}'\rangle^{P}$,
$|\Phi_{5}'\rangle^{S}$, and $|\Phi_{6}'\rangle^{S}$, respectively.
Here,
\begin{equation}\label{eq22}
\begin{split}
|\Phi'_{5}\rangle^{P}=&\frac{1}{2\sqrt{2}}[(|RR\rangle+|LL\rangle)_{AB}
(|RR\rangle-|LL\rangle)_{CD}\\
&-(|RR\rangle-|LL\rangle)_{AB}(|RL\rangle-|LR\rangle)_{CD}],\\
|\Phi'_{6}\rangle^{P}=&\frac{1}{2\sqrt{2}}[(|RL\rangle+|LR\rangle)_{AB}
(|RR\rangle+|LL\rangle)_{CD}\\
&+(|RL\rangle-|LR\rangle)_{AB}(|RL\rangle+|LR\rangle)_{CD}], \\
|\Phi'_{5}\rangle^{S}=&\frac{1}{2\sqrt{2}}[(|a_{1}b_{1}\rangle+|a_{2}b_{2}\rangle)_{AB}
(|c_{1}d_{1}\rangle-|c_{2}d_{2}\rangle)_{CD}\\
&-(|a_{1}b_{1}\rangle-|a_{2}b_{2}\rangle)_{AB}(|c_{1}d_{2}\rangle-|c_{2}d_{1}\rangle)_{CD}],\\
|\Phi'_{6}\rangle^{S}=&\frac{1}{2\sqrt{2}}[(|a_{1}b_{2}\rangle+|a_{2}b_{1}\rangle)_{AB}
(|c_{1}d_{1}\rangle+|c_{2}d_{2}\rangle)_{CD}\\
&+(|a_{1}b_{2}\rangle-|a_{2}b_{1}\rangle)_{AB}(|c_{1}d_{2}\rangle+|c_{2}d_{1}\rangle)_{CD}].
\end{split}
\end{equation}

Finally, photons $C$ and $D$ are detected. If photons $C$ and
$D$ are in the  even-parity state in the polarization DOF (the spatial-mode
DOF), the first step of the hyper-EPP is finished directly. If photons $C$ and $D$ are
in the odd-parity state in the polarization DOF (the spatial-mode DOF),  the first step of the  hyper-EPP is finished after performing a phase-flip operation
$\sigma_{z}^{P}=|R\rangle\langle R|-|L\rangle\langle
L|(\sigma_{z}^{S}=|b_{1}\rangle\langle b_{1}|-|b_{2}\rangle\langle
b_{2}|)$ on photon $B$ in the polarization DOF
(spatial-mode DOF). However, Alice and Bob should discard the photon
pair $AB$ in this case, as the photons $AB$ now are in a mixed state
composed of the correct and bit-flip-error
states with equal probabilities.

In case (3), the states of photons $AC$ and $BD$ are of
different parities in the polarization DOF but of the same parity
in the spatial-mode DOF, which corresponds to the state
$|\phi^{+}\rangle_{AB}^{P}|\psi^{+}\rangle_{CD}^{P}$ or
$|\psi^{+}\rangle_{AB}^{P}|\phi^{+}\rangle_{CD}^{P}$ in the polarization
DOF and the state
$|\phi^{+}\rangle_{AB}^{S}|\phi^{+}\rangle_{CD}^{S}$ or
$|\psi^{+}\rangle_{AB}^{S}|\psi^{+}\rangle_{CD}^{S}$ in the spatial-mode
DOF. For the polarization DOF, the states of the four-photon system $ABCD$
in this case are the same as those in case (2),
leading to the same results as those in case (2) for the polarization DOF. For
the spatial-mode DOF, the states of the four-photon system $ABCD$ in this case are the
same as those in case (1), leading to the same results as those in  case (1) for the spatial-mode DOF.
In this case, Alice and Bob would reserve the photon pair $AB$ and then
perform the second purification step of our hyper-EPP.

In case (4), the states of photons $AC$ and $BD$ are of the same
parity in the polarization DOF but in different parities in the
spatial-mode DOF, which corresponds to the state
$|\phi^{+}\rangle_{AB}^{P}|\phi^{+}\rangle_{CD}^{P}$ or
$|\psi^{+}\rangle_{AB}^{P}|\psi^{+}\rangle_{CD}^{P}$ in the polarization
DOF and the state
$|\phi^{+}\rangle_{AB}^{S}|\psi^{+}\rangle_{CD}^{S}$ or
$|\psi^{+}\rangle_{AB}^{S}|\phi^{+}\rangle_{CD}^{S}$ in the spatial-mode
DOF. For the polarization DOF,  the states of the four-photon system $ABCD$
in this case are the same as those in case (1),
leading to the same results as those in case (1) for the polarization DOF. For
the spatial-mode DOF, the states of the four-photon system $ABCD$ in this case are the
same as those in case (2), leading to the same results as those in  case (2) for the spatial DOF. In this case, Alice and Bob would also reserve the photon pair $AB$ and then perform
the second purification step of our hyper-EPP as well.

\begin{figure}[!ht]
\begin{center}
\includegraphics[width=8.6cm,angle=0]{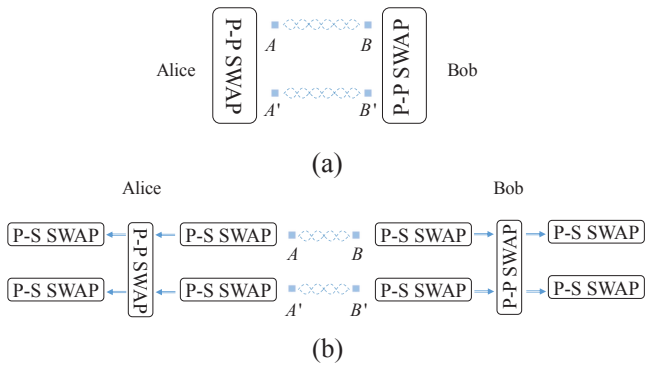} \hspace{30pt}
\caption{Schematic diagram of the second purification step of the hyper-EPP.
(a) Schematic diagram of the first possible purification arrangement to purify the photon pair $AB$. (b) Schematic diagram of the second possible purification arrangement to purify the photon pair $A'B'$. }
\label{fig6}
\end{center}
\end{figure}

\subsection{The second purification step of the hyper-EPP}

For two-photon systems created in cases (3) and  (4), the hyperentangled
photons with a purified higher fidelity
are created in the spatial-mode and polarization DOFs, respectively. One can perform
the second purification step of the hyper-EPP by  pumping the higher-fidelity components into
the same photon pair, leading to a purified photon pair with higher
fidelities in both the polarization and spatial-mode DOFs. This will significantly improve
the efficiency of our hyper-EPP.
The schematic diagram of this procedure is shown in Fig.~\ref{fig6}.
Suppose Alice and Bob share two photon pairs $AB$ and $A'B'$ after the first purification
step. $AB$ is created from case (3) with a higher fidelity in the spatial-mode DOF
and  $A'B'$ is created from case (4) with a higher fidelity in the polarization DOF.

In general, there are two possible purification arrangements. The first one is to purify the photon pair $AB$  with the photon pair $A'B'$.
Alice performs a fidelity-robust P-P SWAP gate on the photons $A$ and $A'$ and Bob performs a
fidelity-robust P-P SWAP gate on the photons $B$ and $B'$ as shown in
Fig.~\ref{fig6}(a). Using these P-P SWAP gates, Alice and Bob complete the purification of the photon pair $AB$, by
replacing the lower-fidelity polarization state of $AB$ with the higher-fidelity polarization state of $A'B'$.
Now, both the polarization and spatial-mode DOFs of photon pair $AB$ are of higher fidelities,
which are, in principle, the same as those in case (1). The other purification arrangement is to purify the photon pair $A'B'$  with the photon pair $AB$.
This arrangement is much similar to the first one while one needs to replace the P-P SWAP gate with a SWAP gate on two photons in the spatial-mode DOF, denoted as the S-S SWAP gate, which can be constructed by sandwiching the P-P SWAP gate with two P-S SWAP gates, shown in Fig.~\ref{fig6}(b). The performances of both purification arrangements are, in principle, the same as each other, since the P-S SWAP gate consists of linear optical elements.

\begin{figure}[th]
\centering
\includegraphics[width=8cm,angle=0]{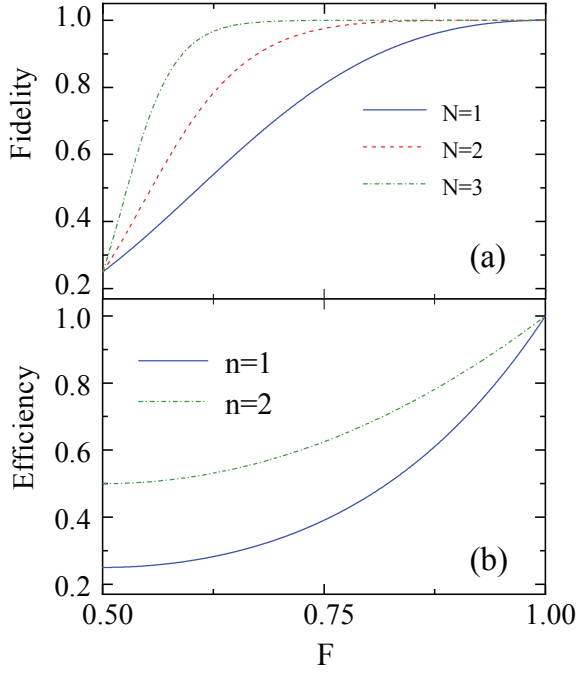}
\caption{(a) Purified fidelity versus the initial fidelity after performing  the $N$-round hyper-EPP.
(b) The efficiency versus the initial fidelity with the hyper-EPP including one purification step  $Y_1$ ($n=1$)
and our improved hyper-EPP including the second purification step $Y_2$ ($n=2$).
} \label{fig7}
\end{figure}

\subsection{Performance of the hyper-EPP}

After these two purification steps, the first round of the hyper-EPP process
is accomplished. Interestingly, the photon pair $AB$ obtained from  case (1)
in the first purification step completes the first round of the hyper-EPP directly. The
photon pair $AB$ obtained by subsequently performing two steps is completed by purifying
the spatial-mode  DOF with the first purification step and purifying
the polarization DOF with the second purification step.
After the first round of hyper-EPP, the purified state of the photon pair $AB$ is changed into
\begin{equation}\label{eq23}
\begin{split}
\rho'_{AB}=&[F'_{1}|\phi^{+}\rangle_{AB}^{P} \langle \phi^{+}|
+(1-F'_{1})|\psi^{+}\rangle_{AB}^{P} \langle \psi^{+}|]\\
&\otimes[F'_{2}|\phi^{+}\rangle_{AB}^{S} \langle \phi^{+}|
+(1-F'_{2})|\psi^{+}\rangle_{AB}^{S} \langle \psi^{+}|],
\end{split}
\end{equation}
where $F'_{i}=\frac{F_{i}^{2}}{F_{i}^{2}+(1-F_{i})^{2}}$ for $(i=1,2)$. The fidelity of the photon pair $AB$ after performing the hyper-EPP once is
$F'=F'_{1}F'_{2}$, which is higher than the initial one
$F=F_{1}F_{2}$ for the case with $F_{i}>1/2$. Furthermore, the fidelity of the
photon pair $AB$ can be improved further by performing multiple-round
hyper-EPP in both the polarization and spatial-mode DOFs. The
purified  fidelities versus the initial fidelities after performing one-, two-, and
three-round hyper-EPP are shown in Fig.~\ref{fig7}(a).

The efficiency of obtaining  a purified photon pair with higher fidelities
in both DOFs after the first purification step is $Y_{1}$, while the efficiency of obtaining a
photon pair with  improved fidelities in both DOFs after introducing the second purification step  is $Y_{2}$,
\begin{equation}\label{eq24}
\begin{split}
Y_{1}=&[F_{1}^{2}+(1-F_{1})^{2}][F_{2}^{2}+(1-F_{2})^{2}],\\
Y_{2}=&[F_{1}^{2}+(1-F_{1})^{2}][F_{2}^{2}+(1-F_{2})^{2}] \\
&+min\{[F_{1}^{2}+(1-F_{1})^{2}][2F_{2}(1-F_{2})],\\
&[2F_{1}(1-F_{1})][F_{2}^{2}+(1-F_{2})^{2}]\}.
\end{split}
\end{equation}
The efficiencies $Y_{1}$ and $Y_{2}$ are shown in Fig.~\ref{fig7}(b)
for the case with $F_{1}=F_{2}=F$. One can easily find that the
efficiency of our hyper-EPP $Y_{2}$ is significantly increased. For the initial fidelity $F<0.75$,
$Y_{2}$ is larger than $2Y_{1}$.

\begin{figure}[!th]
\centering
\includegraphics[ width=8cm,angle=0]{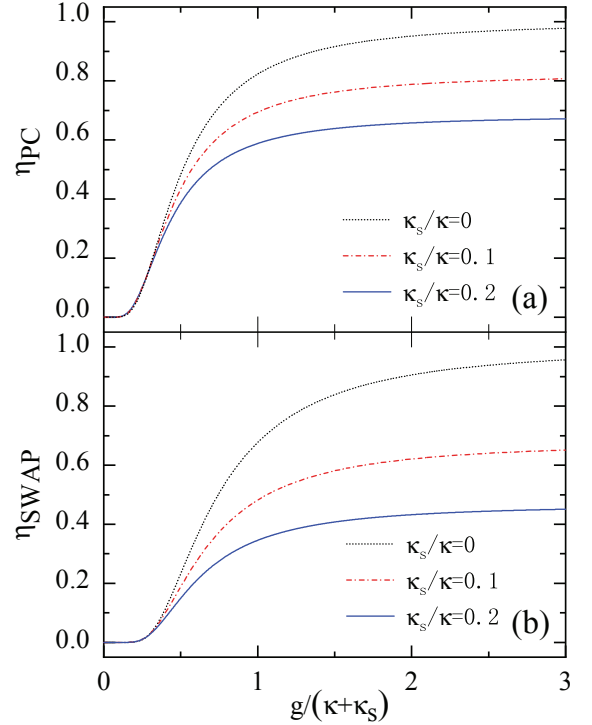}
\caption{(a) The efficiency $\eta_{PC}$ of our fidelity-robust parity-check QNDs in both the
polarization DOF and the spatial-mode DOF for $\kappa_{s}/\kappa$=0.2, $\kappa_{s}/\kappa$=0.1, and $\kappa_{s}/\kappa$=0, which are represented, respectively, by the blue solid line, the red dot-dash line, and the black dot line, with $\omega=\omega_{c}=\omega_{X^{-}}$ and
$\gamma/\kappa=0.1$. (b) The efficiency $\eta_{SWAP}$ of our fidelity-robust P-P SWAP gate for $\kappa_{s}/\kappa$=0.2, $\kappa_{s}/\kappa$=0.1, and $\kappa_{s}/\kappa$=0, which are represented respectively by the blue solid line, the red dot-dash line, and the black dot line, with $\omega=\omega_{c}=\omega_{X^{-}}$ and $\gamma/\kappa=0.1$.} \label{fig8}
\end{figure}

\section{Discussion and summary} \label{sec6}

The fidelity-robust parity-check QNDs and the SWAP gates are two core blocks for implementing the faithful hyper-EPP in both the polarization and spatial-mode DOFs. In the first purification step of our hyper-EPP, the fidelity-robust parity-check QNDs are used to pick out the effective components from the original
mixed state. We transfer the infidelity originating from practical scattering into a heralded failure that clicks the single-photon detectors. The fidelity of the parity-check QND in each DOF, in principle, approaches unity when the parity-check QND succeeds, and it is  not affected by the imperfection involving in the practical scattering. However, the efficiency of the parity-check QND in each DOF $\eta_{PC}=|T|^{4}$, as shown in Fig.~\ref{fig8}(a), is largely constrained by the practical QD-cavity union. For a typical dipole decay $\gamma/\kappa=0.1$ and cavity side-leakage rate $\kappa_{s}/\kappa$=0.2, the efficiency $\eta_{PC}>65.8\%$ when $g/(\kappa+\kappa_{s})>{2.0}$ and $\omega=\omega_{c}=\omega_{X^{-}}$. Furthermore, as shown in Fig.~\ref{fig8}(a),  the efficiency $\eta_{PC}$ could be further increased by increasing the effective QD-cavity coupling  $g/(\kappa+\kappa_{s})$ and decreasing the side leakage $\kappa_{s}/\kappa$ of the cavity.

In the second purification step of our hyper-EPP, the  SWAP gates, including the P-P SWAP gate and the P-S SWAP gate, work in a controllable arrangement and pump the purified higher-fidelity DOFs of different photon pairs into the same photon pair, which significantly increases the efficiency of our hyper-EPP. Similarly, the infidelity of the practical scattering is also transferred into a heralded failure, which makes the fidelity of our P-P SWAP gate near unity and be robust to the deviation from the ideal single-photon scattering. The efficiency of our P-P SWAP gate $\eta_{SWAP}=|T|^{8}$, as shown in Fig.~\ref{fig8}(b). With the same experimental parameters $\gamma/\kappa=0.1$ and $\kappa_{s}/\kappa$=0.2, we find that $\eta_{SWAP}>43.2\%$ is achieved with $g/(\kappa+\kappa_{s})>{2.0}$ and $\omega=\omega_{c}=\omega_{X^{-}}$.  Similarly, $\eta_{SWAP}$ could also be increased by increasing the effective QD-cavity coupling and decreasing the side leakage of the cavity, as shown in Fig.~\ref{fig8}(b).

Up to now,  we  have  assumed  that single-photon detectors in our quantum circuits work  perfectly with unity efficiency. In the process of constructing  parity-check QNDs and  SWAP gates, the click of a single-photon detector in any modified QD-cavity union is considered inconclusive, which heralds the failure of the corresponding process, but it is also a signal to discard postprocessing of the present hyper-EPP.  Conversely, the no click of the single-photon detector, in fact, projects the hybrid system, consisting of a single photon and a QD, into a superposition state
$\vert \psi_h\rangle=p_0\vert \psi_{loss}\rangle+p_1\vert \psi_{ideal}\rangle$, where $\vert \psi_{ideal}\rangle$ represents the desired output state and $\vert \psi_{loss}\rangle$ represents an error state with a photon loss. The photon loss mainly originates from the decay involving in the nondeterministic scattering by the QD-cavity union. However, when single-photon detectors of a finite efficiency $\eta_d$ are used, a failure click of the detectors can be viewed as an additional photon loss, and it will not change the desired state $p_1\vert \psi_{ideal}\rangle$. Therefore, the finite efficiency of single-photon detectors has no effect on the efficiencies and the fidelities of parity-check QNDs and SWAP gates. In each of the two purification steps, single-photon detectors are used to perform the projection measurements on two auxiliary photons, and if there are no coincident clicks of two single-photon detectors, our hyper-EPP cannot be deemed to be accomplished. In this respect, the inefficiency of single-photon detectors  will decrease the efficiency of our hyper-EPP by a scale of $\eta^2_d$, but it will not decrease the fidelity of our hyper-EPP.

Our faithful hyper-EPP is assisted by the QD embedded in a double-sided optical microcavity, which is an attractive single-photon single-atom interface \cite{QD001,QD002}. The techniques of fast preparing the superposition state  \cite{QD1,QD2}, manipulating \cite{QD3,QD4,QD5,QD6}, and detecting of the QD-confined electron spin \cite{QD7} have been realized experimentally.  Although our parity-check QNDs  and P-P SWAP gate, in principle, work with unity fidelity and are immune to the imperfection involved in a practical scattering process when the ideal QD is used, their fidelities will decrease a little when the realistic QD is used. The electron spin decoherence would decrease the fidelities by the amount of $F=[1+exp(-\Delta t/T_{2}^{e})]/2$ \cite{Hu2}. $\Delta t$ is the time consumed for completing the parity-check QNDs or the P-P SWAP gate, which is about several nanoseconds, while $T_{2}^{e}$ is the electron spin coherence time that is about several microseconds \cite{QDex1}. Therefore, the infidelity due to the finite coherence time of QDs will be less than  $1\%$. In addition, the hole mixing in  a realistic QD will affect the optical selection rule, and then affect the fidelities of our proposal as well \cite{QDex4}. Fortunately, this influence can be efficiently suppressed by engineering the shape and size of QDs or by using different types of QDs \cite{QDmixhole}.

In summary, we have proposed a faithful hyper-EPP for two-photon
systems in both the polarization and spatial-mode DOFs. By introducing an additional swapping operation between different photon pairs, the parties in quantum communication can exploit the original abandoned photon pairs in the first purification step  to distil high-fidelity ones with the second purification step, which significantly increases the efficiency of our hyper-EPP. The fidelity-robust parity-check QNDs and SWAP gates used in the first and second purification steps are constructed in a new structure and their fidelities are robust to the practical deviations from the ideal single-photon scattering. The imperfections, such as the finite side leakage of the microcavity and the finite coupling between the QD and the cavity mode, will only decrease the efficiencies in a
heralded way but not decrease the fidelities. This makes our hyper-EPP works faithfully for high-capacity quantum networks.  Furthermore, our faithful hyper-EPP could be extended directly to purify the photon systems entangled in the single polarization or spatial-mode DOF by simplifying our fidelity-robust parity-check QNDs to that for the corresponding DOF, and it can also be extended to the case for purifying the photon systems hyperentangled in the polarization and  multiple-spatial-mode DOFs by generalizing our parity-check QNDs and SWAP gates in a similar way. We believe our hyper-EPP will contribute significantly to the development of high-capacity quantum communication and quantum networks in the future.

\section*{Acknowledgments}

This work is supported by the National Natural Science Foundation
of China under Grants No. 11674033, No.  11474026, and No. 11505007, and the Natural Science Foundation of Jiangsu Province under Grant No. BK20180461.



\begin{thebibliography}{99}

\bibitem{tele1}C. H. Bennett, G. Brassard, C. Crepeau, R. Jozsa,
A. Peres, and W. K. Wootters, Teleporting an unknown quantum state via
dual classical and Einstein-Podolsky-Rosen channels, Phys. Rev. Lett. \textbf{70}, 1895 (1993).


\bibitem{dense1} C. H. Bennett and S. J. Wiesner,
Communication via one- and two-particle operators on Einstein-Podolsky-Rosen states,
Phys. Rev. Lett. \textbf{69}, 2881--2884 (1992).


\bibitem{super2}X. S. Liu, G. L. Long, D. M. Tong, and L. Feng,
General scheme for superdense coding between multiparties, Phys. Rev. A \textbf{65}, 022304 (2002).


\bibitem{Ekert} A. K. Ekert, Quantum cryptography based on Bell's
theorem, Phys. Rev. Lett. \textbf{67}, 661--663 (1991).


\bibitem{BBM92} C. H. Bennett, G. Brassard, and N. D. Mermin,
Quantum cryptography without Bell's theorem, Phys. Rev. Lett.
\textbf{68}, 557--559 (1992).


\bibitem{QSS1} M. Hillery, V. Bu\v{z}ek, and A. Berthiaume,
Quantum secret sharing, Phys. Rev. A \textbf{59}, 1829 (1999).


\bibitem{QSDC1}G. L. Long and X. S. Liu,
Theoretically efficient high-capacity quantum-key-distribution scheme,
Phys. Rev. A \textbf{65}, 032302 (2002).

\bibitem{twostep} F. G. Deng, G. L. Long, and X. S. Liu,
Two-step quantum direct communication protocol using the
Einstein-Podolsky-Rosen pair block,  Phys. Rev. A \textbf{68},
042317 (2003).

\bibitem{twostepexp} W. Zhang, D. S. Ding, Y. B. Sheng, L. Zhou, B. S. Shi, and G. C.
Guo, Quantum Secure Direct Communication with Quantum Memory, Phys.
Rev. Lett. \textbf{118}, 220501 (2017).

\bibitem{hyper0} P. G. Kwiat, Hyper-entangled states, J. Mod. Opt. \textbf{44}, 2173 (1997).




\bibitem{super1}J. T. Barreiro, T. C. Wei, and P. G. Kwiat, Beating the channel capacity limit for linear photonic superdense coding, Nat. Phys. \textbf{4}, 282 (2008).

\bibitem{BSA1}P. G. Kwiat and H. Weinfurter, Embedded Bell-state analysis,
Phys. Rev. A \textbf{58}, R2623--R2626 (1998).

\bibitem{BSA2}S. P. Walborn, S. P\'{a}dua, and C. H. Monken,
Hyperentanglement-assisted Bell-state analysis, Phys. Rev. A \textbf{68}, 042313 (2003).

\bibitem{BSA3}C. Schuck, G. Huber, C. Kurtsiefer, and H. Weinfurter,
Complete deterministic linear optics Bell state analysis,
Phys. Rev. Lett. \textbf{96}, 190501 (2006).

\bibitem{BSA4}M. Barbieri, G. Vallone, P. Mataloni, and F. De Martini,
Complete and detrministic deicrimination of polarization Bell states
assisted by momentum entanglement, Phys. Rev. A \textbf{75}, 042317 (2007).

\bibitem{EPP6}Y. B. Sheng and F. G. Deng, Deterministic entanglement
purification and complete nonlocal Bell-state analysis with hyperentangledment,
Phys. Rev. A \textbf{81}, 032307 (2010).

\bibitem{EPP7}Y. B. Sheng and F. G. Deng,
One-step deterministic polarizationentanglement purification using spatial entanglement,
Phys. Rev. A \textbf{82}, 044305 (2010).

\bibitem{EPP8}X. H. Li, Deterministic polarization-entanglement
purification using spatial entanglement, Phys. Rev. A \textbf{82}, 044304 (2010).

\bibitem{EPP9}F. G. Deng,
One-step error correction for multipartite polarization entanglement,
Phys. Rev. A \textbf{83}, 062316 (2011).

\bibitem{EPP10}Y. B. Sheng and L. Zhou,
Deterministic polarization entanglement purification using time-bin entanglement,
Laser Phys. Lett. \textbf{11}, 085203 (2014).

\bibitem{cluster}G. Vallone, E. Pomarico, P. Mataloni, F. De Martini, and V. Berardi, Realization and characterization of a two-photon four-qubit linear cluster state, Phys. Rev. Lett. \textbf{98}, 180502 (2007).

\bibitem{oneway1} H. J. Briegel and R. Raussendorf, Persistent Entanglement in Arrays of Interacting Particles, Phys. Rev. Lett. \textbf{86}, 910 (2001).

\bibitem{oneway2} K. Chen, C. M. Li, Q. Zhang, Y. A. Chen, A. Goebel, S. Chen, A. Mair, and J. W. Pan, Experimental Realization of One-Way Quantum Computing with Two-Photon Four-Qubit Cluster States, Phys. Rev. Lett. \textbf{99}, 120503 (2007).

\bibitem{oneway3} G. Vallone, E. Pomarico, F. De Martini, and P. Mataloni, Active One-Way Quantum Computation with Two-Photon Four-Qubit Cluster States, Phys. Rev. Lett. \textbf{100}, 160502 (2008).

\bibitem{hyperQC1}
 B. C. Ren, H. R. Wei, and F. G. Deng, Deterministic photonic  spatial-polarization  hyper-controlled-not  gate  assisted  by  a  quantum dot  inside  a  one-side  optical  microcavity, Laser Phys. Lett. \textbf{10}, 095202 (2013).

\bibitem{hyperQC2} B. C. Ren and F.G. Deng, Hyper-parallel photonic quantum computing with coupled quantum dots, Sci. Rep. \textbf{4}, 4623 (2014).

\bibitem{hyperQC3}B. C. Ren, G. Y. Wang, and F. G. Deng, Universal hyperparallel hybrid photonic quantum gates with dipole-induced transparency in the weak-coupling regime, Phys. Rev. A \textbf{91}, 032328 (2015).



\bibitem{hyperQC4}
T.  J.  Wang,  Y.  Zhang,  and  C.  Wang,  Universal  hybrid hyper-controlled quantum gates assisted  by quantum  dots  in  optical  double-sided  microcavities,  Laser
Phys. Lett. \textbf{11}, 025203 (2014).

\bibitem{hyperQC5}
T.  Li  and  G.  L.  Long,  Hyperparallel  optical  quantum
computation assisted by atomic ensembles embedded in
double-sided optical cavities,  Phys. Rev. A \textbf{94},  022343
(2016).


\bibitem{hyperQC6} B. C. Ren and F. G. Deng, Robust hyperparallel photonic quantum entangling gate with cavity QED, Opt. Express \textbf{25}, 10863 (2017).

\bibitem{hyperQC7}G. Y. Wang, T. Li, Q. Ai, and F. G. Deng, Self-error-corrected hyperparallel photonic quantum computation working with both the polarization and the spatial-mode degrees of freedom, Opt. Express \textbf{26}, 23333 (2018).








\bibitem{hyper1} J. T. Barreiro, N. K. Langford, N. A. Peters, and P. G. Kwiat,
Generation of hyperentangled photon pairs, Phys. Rev. Lett. \textbf{95}, 260501 (2005).

\bibitem{hyper2} C. Cinelli, M. Barbieri, F. De Martini, and P. Mataloni,
Realization of hyperentangled two-photon states, Laser Phys. \textbf{15}, 124 (2005).

\bibitem{hyper3} M. Barbieri, C. Cinelli, F. De Martini, and P. Mataloni,
Generation of (2$\times$2) and (4$\times$4) two-photon states with tunable
degree of entanglement and mixedness, Fortschr. Phys. \textbf{52}, 1102 (2004).

\bibitem{hyper4} M. Barbieri, C. Cinelli, P. Mataloni, and F. De Martini,
Polarization-momentum hyperentangled states: Realization and characterization,
Phys. Rev. A \textbf{72}, 052110 (2005).

\bibitem{hyper5} D. Bhatti, J. von Zanthier, and G. S. Agarwal,
Entanglement of polarization and orbital angular momentum,
Phys. Rev. A \textbf{91}, 062303 (2015).

\bibitem{hyper6} A. Rossi, G. Vallone, A. Chiuri, F. De Martini, and P. Mataloni,
Mulipath Entanglement of Two Photons, Phys. Rev. Lett. \textbf{102}, 153902 (2009).

\bibitem{hyper7} W. B. Gao, C. Y. Lu, X. C. Yao, P. Xu, O. Guhne, A. Goebel, Y. A. Chen,
C. Z. Peng, Z. B. Chen, and J. W. Pan,
Experimental demonstration of a hyper-entangled ten-qubit Schr\"{o}dinger cat state,
Nat. Phys. \textbf{6}, 331 (2010).

\bibitem{hyper8} G. Vallone, R. Ceccarelli, F. De Martini, and P. Mataloni,
Hyperentanglement of two photons in three degrees of freedom,
Phys. Rev. A \textbf{79}, 030301(R) (2009).

\bibitem{repeater1} H. J. Briegel, W. D\"{u}r, J. I. Cirac, and P. Zoller, Quantum repeaters: the role of imperfect local operations in quantum communication, Phys. Rev. Lett. \textbf{81}, 5932--5935 (1998).

\bibitem{repeater2} W. D\"{u}r, H. J. Briegel, J. I. Cirac, and P. Zoller, Quantum repeaters based on entanglement purification, Phys. Rev. A \textbf{59}, 169--181 (1999).
\bibitem{repeater3} L. M. Duan, M. D. Lukin, J. I. Cirac, and P. Zoller, Long-distance quantum communication
with atomic ensembles and linear optics, Nature (London) \textbf{414}, 413 (2001).

\bibitem{EPP2017}
N. Kalb, A. A. Reiserer, P. C. Humphreys, J. J. W. Bakermans, S. J. Kamerling, N. H. Nickerson, S. C. Benjamin, D. J. Twitchen, M. Markham, and R. Hanson, Entanglement distillation between solid-state quantum network nodes, Science \textbf{356}, 928 (2017).

\bibitem{EPP1} C. H. Bennett, G. Brassard, S. Popescu, B. Schumacher,
J. A. Smolin, and W. K.Wootters, Purification of noise entanglement and
faithful teleportation via noisy channels, Phys. Rev. Lett. \textbf{76}, 722 (1996).


\bibitem{EPP2} D. Deutsch, A. Ekert,R. Jozsa, C. Macchiavello, S. Popescu, and A. Sanpera,
Quantum privacy amplification and the security of quantum cryptography over noisy channel,
Phys. Rev. Lett. \textbf{77}, 2818 (1996).


\bibitem{EPP3}J. W. Pan, C. Simon, \v{C}. Brukner, and A. Zeilinger,
Entanglement purification for quantum communication, Nature (London) \textbf{410}, 10
    67 (2001); J. W. Pan, S. Gasparoni, R. Ursin, G. Weihs, and A. Zeilinger,
    Experimental entanglement purification of arbitrary unknown states,
    Nature (London) \textbf{423}, 417 (2003).


\bibitem{EPP4}C. Simon and J. W. Pan,
Polarization Entanglement Purification using Spatial Entanglement,
Phys. Rev. Lett. \textbf{89}, 257901 (2002).

\bibitem{EPP5}Y. B. Sheng, F. G. Deng, and H. Y. Zhou,
Efficient polarizationentanglement purification based on parametric
down-conversion sources with cross-Kerr nonlinearity, Phys. Rev. A \textbf{77}, 042308 (2008).



\bibitem{hyEPP1}B. C. Ren and F. G. Deng,
Hyperentanglement purification and concentration assisted by diamond NV centers inside
photoic crystal cavities, Laser Phys. Lett. \textbf{10}, 115201 (2013).

\bibitem{hyEPP2}B. C. Ren, F. F. Du, and F. G. Deng,
Two-step hyperentanglement purification with the quantum-state-joining method,
Phys. Rev. A \textbf{90}, 052309 (2014).

\bibitem{hyEPP3}G. Y. Wang, Q. Liu, and F. G. Deng,
Hyperentanglement purification for two-photon six-qubit quantum systems,
Phys. Rev. A \textbf{94}, 032319 (2016).







\bibitem{QD001}
I. Buluta, S. Ashhab, and F. Nori, Natural and artificial atoms for quantum computation, Rep. Prog. Phys. \textbf{74}, 104401  (2011).

\bibitem{QD002}
P. Lodahl, S. Mahmoodian, and S. Stobbe, Interfacing single photons and single quantum dots with photonic nanostructures, Rev. Mod. Phys. 87,
\textbf{347} (2015).



\bibitem{Walls}D. F. Walls and G. J. Milburn, Quantum Optics,
(Springer-Verlag, Berlin, 1994).

\bibitem{Hu1}C. Y. Hu, W. J. Munro, J. L. O'Brien, and J. G. Rarity,
Proposed entanglement beam splitter using a quantumdot spin in a double-sided optical microcavity, Phys. Rev. B \textbf{80}, 205326 (2009).

\bibitem{chara1} G. Vallone, R. Ceccarelli, F. De Martini, and P. Mataloni, Hyperentanglement witness, Phys. Rev. A \textbf{78}, 062305 (2008).



\bibitem{QD1} M. Atat\"{u}re, J. Dreiser, A. Badolato, A. H\"{o}gele, K. Karrai,
and A. Imamoglu, Quantum-dot spin-state preparation with near-unity fidelity,
Science \textbf{312}, 551-553 (2006).




\bibitem{QD2} M. Atat\"{u}re, J. Dreiser, A. Badolato, and A. Imamoglu,
Observation of faraday rotation from a single confined spin, Nat.
Phys. \textbf{3}, 101-106 (2007).


\bibitem{QD3}J. Berezovsky, M. H. Mikkelsen, N. G. Stoltz, L. A. Coldren, and D. D. Awschalom,
Picosecond coherent optical manipulation of a single electron spin in a quantum dot,
Science \textbf{320}, 349-352 (2008).


\bibitem{QD4}D. Press, T. D. Ladd, B. Y. Zhang, and Y. Yamamoto,
Complete quantum control of a single quantum dot spin using ultrafast optical pulses,
Nature (London) \textbf{456}, 218-221 (2008).


\bibitem{QD5}J. A. Gupta, R. Knobel, N. Samarth, and D. D. Awschalom,
Ultrafast manipulation of electron spin coherence, Science \textbf{292}, 2458-2461 (2001).


\bibitem{QD6}P. C. Chen, C. Piermarocchi, L. J. Sham, D. Gammon, and D. G. Steel,
Theory of quantum optical control of a single spin in a quantum dot,
Phys. Rev. B \textbf{69}, 075320 (2004).


\bibitem{QD7}R. Hanson, L. H. Willems van Beveren, I. T. Vink, J. M. Elzerman,
W. J. M. Naber, F. H. L. Koppens, L. P. Kouwenhoven, and L. M. K. Vandersypen,
Single-shot readout of electron spin states in a quantum dot using spin-dependent tunnel rates,
Phys. Rev. Lett. \textbf{94}, 196802 (2005).

\bibitem{Hu2} C. Y. Hu and J. G. Rarity, Loss-resistant state teleportation and entanglement swapping using a quantum-dot spin in an optical microcavity, Phys. Rev. B \textbf{83}, 115303 (2011).


\bibitem{QDex1}A. Greilich, D. R. Yakovlev, A. Shabaev, A. L. Efros, I. A. Yugova,
R. Oulton, V. Stavarache, D. Reuter, A. Wieck, and M. Bayer,
Mode locking of electron spin coherences in singly charged quantum dots,
Science \textbf{313}, 341-345 (2006).




\bibitem{QDex4}G. Bester, S. Nair, and A. Zunger,
Pseudopotential calculation of the excitonic fine structure of
million-atom self-assembled In$_{1-x}$Ga$_x$As/GaAs quantum dots,
Phys. Rev. B \textbf{67}, 161306 (2003).

\bibitem{QDmixhole}
R. B. Liu, W. Yao, and L. J. Sham, Quantum computing by optical control of electron spins,
 Adv. Phys. 59, \textbf{703} (2010).





\end{thebibliography}
\end{document}